\documentclass[aps,twocolumn,prd,superscriptaddress,nofootinbib,floatfix]{revtex4-1}

\usepackage{mathtools}
\usepackage{amsfonts}
\usepackage{amssymb}
\usepackage{mathrsfs}
\usepackage{bbm}
\usepackage{slashed}
\usepackage{amsmath}
\usepackage{bm}
\usepackage{bbold}

\usepackage{graphicx}
\usepackage{color}
\usepackage{colortbl}
\usepackage{array}
\usepackage[dvipsnames]{xcolor}
\usepackage{xcolor}

\usepackage{float}
\usepackage{placeins}
\usepackage{booktabs}
\usepackage[caption=false]{subfig}
\usepackage{makecell}
\usepackage{tabstackengine}

\usepackage{xspace}
\usepackage{hyperref}
\usepackage[nameinlink]{cleveref}
\usepackage{bookmark}
\usepackage{siunitx}

\usepackage{xifthen}
\usepackage{xcolor}
\hypersetup{
	colorlinks,
	linkcolor={red!75!black},
	citecolor={blue!75!black},
	urlcolor={blue!75!black}
}

\usepackage[utf8]{inputenc}
\allowdisplaybreaks[4]

\newcommand{\feyn}[1]{
	\setbox0=\hbox{\ensuremath{#1}}
	\hbox to\wd0{\hbox to0pt{\hbox to\wd0{\hss/\hss}\hss}\box0}}

\setkeys{Gin}{width=0.48\textwidth}
\graphicspath{{./figures/}}

\definecolor{red}{rgb}{1,0,0}

\def\be{\begin{equation}}
\def\ee{\end{equation}}
\def\bea{\begin{eqnarray}}
\def\eea{\end{eqnarray}}

\newcommand{\gettitle}{}

\hypersetup{
	colorlinks,
	linkcolor={red!75!black},
	citecolor={blue!75!black},
	urlcolor={blue!75!black},
	pdftitle={\gettitle},
	pdfauthor={Zelle},
	pdfkeywords={effective theory} {analytic continuation}
	{correlations functions} {hadronization}
	{functional renormalisation group}
	{real time} {bound states}
	bookmarksopen=true,
	bookmarksopenlevel=2,
	bookmarksnumbered=true
}

\begin{document}

\title{Toward precise $\xi$ gauge fixing for the lattice QCD}

\collaboration{\bf{$\chi$QCD Collaboration}}

\author{
\includegraphics[width=0.12\linewidth]{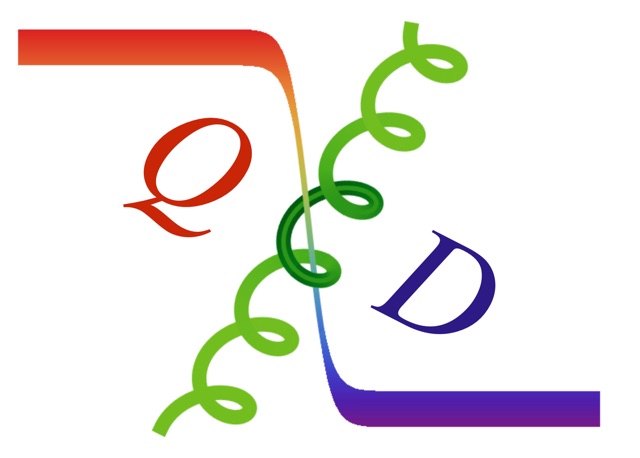}\\
{\bf ($\chi$QCD Collaboration)}\\
Li-Jun Zhou}
\affiliation{School of Physics, Dalian University of Technology, Dalian, 116024, P.R. China}

\author{Dian-Jun Zhao}
\affiliation{School of Science and Engineering, The Chinese University of Hong Kong, Shenzhen 518172, China}

\author{Wei-jie Fu}
\email{Corresponding author: wjfu@dlut.edu.cn}
\affiliation{School of Physics, Dalian University of Technology, Dalian, 116024, P.R. China}	

\author{Chun-Jiang Shi}
\affiliation{Institute of High Energy Physics, Chinese Academy of Sciences, Beijing 100049, China}

\author{Ji-Hao Wang}
\affiliation{University of Chinese Academy of Sciences, School of Physical Sciences, Beijing 100049, China}
\affiliation{CAS Key Laboratory of Theoretical Physics, Institute of Theoretical Physics, Chinese Academy of Sciences, Beijing 100190, China}

\author{Yi-Bo Yang}
\email{Corresponding author: ybyang@itp.ac.cn}
\affiliation{University of Chinese Academy of Sciences, School of Physical Sciences, Beijing 100049, China}
\affiliation{CAS Key Laboratory of Theoretical Physics, Institute of Theoretical Physics, Chinese Academy of Sciences, Beijing 100190, China}
\affiliation{School of Fundamental Physics and Mathematical Sciences, Hangzhou Institute for Advanced Study, UCAS, Hangzhou 310024, China}
\affiliation{International Centre for Theoretical Physics Asia-Pacific, Beijing/Hangzhou, China}

\date{\today}

\begin{abstract}
Lattice QCD provides a first-principles framework for solving Quantum Chromodynamics (QCD). However, its application to off-shell partons has been largely restricted to the Landau gauge, as achieving high-precision $\xi$-gauge fixing on the lattice poses significant challenges. Motivated by a universal power-law dependence of off-shell parton matrix elements on gauge-fixing precision in the Landau gauge, we propose an empirical precision extrapolation method to approximate high-precision $\xi$-gauge fixing. By properly defining the bare gauge coupling and then the effective $\xi$, we validate our $\xi$-gauge fixing procedure by successfully reproducing the $\xi$-dependent RI/MOM renormalization constants for local quark bilinear operators at 0.3\% level, up to $\xi \sim 1$.
\end{abstract}

\maketitle

\section{Introduction}

The gauge invariance is a crucial property of gauge theories. One of the most fundamental principles in constructing a gauge theory is that the Lagrangian should remain gauge-invariant. But the usual quantization and also perturbative calculation requires to introduce an additional gauge fixing term, e.g., $(\partial_{\mu}A^{\mu})^2/(2\xi)$,  while the choice of gauge can be rather arbitrary and may vary among different physicists. For example, the Feynman gauge with $\xi=1$ simplify the form of the gauge boson propagator and then all the perturbative calculation, while the Landau gauge with $\xi=0$ can minimize the loop correction of the quark field at the 1-loop level. Nevertheless, physical observables derived from different gauge fixing choices are also gauge-invariant as the experimental value is evidently unique. 

The scenario in lattice QCD differs slightly from the framework described above. In lattice QCD, the fundamental gauge degree of freedom--the gauge potential \( A_{\mu}(x) \) at a given spacetime point \( x \)--is replaced by the gauge link $U_\mu(x+\hat{\mu}a/2) = e^{ i a g \int_{x}^{x+\hat{\mu}a} \mathrm{d}y^\mu A_\mu(y) }$ connecting \( x \) and \( x + \hat{\mu}a \). Gauge invariance is then automatically ensured as long as the gauge links terminate at quark fields or form closed loops, as in hadronic correlation functions.  

On the other hand, gauge-dependent quantities--such as off-shell parton (quark and gluon) propagators and interaction vertices--vanish entirely unless gauge fixing is imposed. Those quantities are important ingredients of nonperturbative functional QCD, such as the Dyson-Swinger Equations (DSE)~\cite{Chang:2010hb, Qin:2010nq, Bashir:2012fs, Fischer:2014ata, Gao:2015kea, Aguilar:2019teb, Gao:2020fbl, Roberts:2021nhw,Gunkel:2021oya, Chang:2021vvx} and the functional renormalization group (fRG)~\cite{Mitter:2014wpa, Braun:2014ata, Cyrol:2016tym, Cyrol:2017ewj, Fu:2019hdw, Fu:2022uow, Fu:2024ysj, Ihssen:2024miv, Fu:2025hcm, Zhang:2025ofc, Dupuis:2020fhh, Fu:2022gou}. Usually the gauge-dependent propagators and vertices are computed in functional QCD with the Landau gauge $\xi=0$, from which benchmark comparison between the functional QCD and lattice QCD can be made, see e.g.~\cite{Fu:2025hcm} for a recent study. Lattice QCD simulations at low energy scales with the Landau gauge have shown kinds of highly-nontrivial features at hadron scale, such as the emergent masses of the quark and gluon~\cite{Bowman:2005vx, Boucaud:2018xup, Chang:2021vvx}, non-degenerate gauge coupling from the gluon-ghost~\cite{Zafeiropoulos:2019flq} and triple gluon vertices~\cite{Aguilar:2019uob}, and so on, which are consistent with the results of functional QCD, cf. e.g.,~\cite{Mitter:2014wpa, Braun:2014ata, Cyrol:2016tym, Cyrol:2017ewj, Fu:2019hdw, Ihssen:2024miv, Fu:2025hcm}. However, most of those lattice calculations are restricted to the Landau gauge, despite the fact that lattice implementations of general \( \xi \)-gauge fixing~\cite{Fujikawa:1972fe} were proposed years ago~\cite{Giusti:1996kf,Cucchieri:2009kk,Bicudo:2015rma}. The primary challenge lies in severe convergence issues that arise at large \( \xi \) and/or strong gauge coupling \( g \), making the application of \( \xi \)-gauge fixing to realistic configurations numerically demanding, to understand how those inferred features of parton are sensitive to the specific Landau gauge fixing.  

Recently, the dependence of gauge links and non-local operators on gauge-fixing precision has been investigated at multiple lattice spacings and for varying gauge link lengths, in both Landau and $\xi$ gauges~\cite{Zhang:2024omt}. The values of these quantities follow an empirical power law in terms of gauge-fixing precision, regardless of the gauge link length. In this work, we further validate that this power law also holds for local operators with different gamma matrices and off-shell momenta. Based on this, we propose a precision-extrapolation method to approximate high-precision $\xi$-gauge fixing with controllable systematic uncertainty.

The paper is organized as follows: In Sec.~\ref{sec:theofra}, we briefly review the gauge-fixing procedure on the lattice for Landau and $\xi-$gauge, and also the idea of the precision extrapolation based on the empirical power law observed in Ref.~\cite{Zhang:2024omt}. Further numerical evidence of the precision extrapolation in the Landau gauge, are presented in Sec.~\ref{sec:precextralocal}. Section~\ref{sec:xires} presents our results on the non-perturbative $\xi$-gauge dependence of quark bilinear operators, including a detailed comparison with perturbative calculations. Finally, Sec.~\ref{sec:sum} provides a concise summary of our findings.  

\section{Methodology}\label{sec:theofra}

In the path integral formalism, gauge fixing with the additional Lagrangian term $(\partial_{\mu}A^{\mu})^2/(2\xi)$ can be equivalently implemented by introducing $3^2-1$ random variables $\Lambda^a$ which follow the distribution $P(\Lambda^a(x))=\frac{1}{\sqrt{2\pi \xi}}\exp\left\{-\frac{1}{2\xi}[\Lambda^a(x)]^2\right\}$. The gauge fixing condition is then enforced by integrating over $\Lambda \equiv \Lambda^a t^a$ with the delta function constraint $\delta\left(\partial_{\mu}A^{\mu}(x) - \Lambda(x)\right)$, where $t^a$ are the generators of the adjoint representation of $\mathrm{SU}(3)$.

On the lattice, the delta function constraint is discretized into the gauge-fixing condition:  
\begin{align} \label{eq:condition}
\Delta(x) &\equiv \sum_{\mu, \eta = \pm} \eta \left[ \frac{U_\mu(x + \eta \frac{\hat{\mu}}{2} a) - U_\mu^\dagger(x + \eta \frac{\hat{\mu}}{2} a)}{2 {\rm i} g_0} \right]_{\text{Traceless}}\nonumber\\
&\quad\quad - \Lambda(x)a^2 = 0,  
\end{align}  
and the integration over $\Lambda$ can be efficiently performed by averaging over different gauge configurations with independent $\Lambda$. The bare gauge coupling \( g_0 \) in Eq.~(\ref{eq:condition}) can be defined in multiple ways, differing at next-to-leading order in \( \alpha_s \). As an example, consider the tadpole-improved tree-level Symanzik gauge action \( S_g \), defined as:  

\[
S_g = \frac{1}{3} \mathrm{Re} \sum_{x, \mu < \nu} \mathrm{Tr} \left[ 1 - \beta \left( \frac{5}{3} \mathcal{P}^U_{\mu \nu}(x) - \frac{\mathcal{R}^U_{\mu \nu}(x)}{12 u_0^2} \right) \right],
\]  
where  
\begin{align}  
\mathcal{P}^U_{\mu \nu}(x) &= U_\mu(x+\frac{a}{2}\hat{\mu}) U_\nu(x + a \hat{\mu}+\frac{a}{2}\hat{\nu}) \nonumber\\
&\quad\quad\times U^\dagger_\mu(x +\frac{a}{2}\hat{\mu}+ a \hat{\nu}) U^\dagger_\nu(x+\frac{a}{2}\hat{\nu}), \nonumber \\  
\mathcal{R}^U_{\mu \nu}(x) &= U_\mu(x+\frac{a}{2}\hat{\mu}) U_\mu(x + \frac{3a}{2}\hat{\mu}) \nonumber\\
&\quad\quad\times U_\nu(x + 2a \hat{\mu}+\frac{a}{2}\hat{\nu})   
 U^\dagger_\mu(x + \frac{3a}{2} \hat{\mu} + a \hat{\nu}) \nonumber \\
 &\quad\quad\times  U^\dagger_\mu(x + \frac{a}{2} \hat{\mu} + a \hat{\nu}) U^\dagger_\nu(x+\frac{a}{2} \hat{\mu}),  
\end{align}  
and the tadpole improvement factor $u_0$ is given by:  
 $u_0 = \langle \mathrm{Re} \mathrm{Tr} \sum_{x, \mu < \nu} \mathcal{P}^{U}_{\mu \nu}(x)/(6 N_c V)\rangle^{1/4}$.

Then we can have three definitions of $g_0$ in Eq.~(\ref{eq:condition}):

1) Naive definition: $g^{(a)}_0=\sqrt{6/\beta}$;

2) Full tadpole improvement:  Including \( u_0 \) in both the action and also gauge link in the gauge fixing condition, and then Eq.~(\ref{eq:condition}) should be rewritten into 
\begin{align} \label{eq:condition_2}
&\Lambda(x)a^2\nonumber\\
&=\sum_{\mu, \eta = \pm} \eta \left[ \frac{\frac{U_\mu(x + \eta \frac{\hat{\mu}}{2} a)}{u_0} - \frac{U_\mu^\dagger(x + \eta \frac{\hat{\mu}}{2} a)}{u_0} }{2 {\rm i} \sqrt{6/\beta/u_0^4}} \right]_{\text{Traceless}} \nonumber\\
&=\sum_{\mu, \eta = \pm} \eta \left[ \frac{U_\mu(x + \eta \frac{\hat{\mu}}{2} a) - U_\mu^\dagger(x + \eta \frac{\hat{\mu}}{2} a)}{2 {\rm i} \sqrt{6/\beta}/u_0} \right]_{\text{Traceless}}.
\end{align} 
Thus it leads to a effective gauge coupling $g^{(b)}_0=\sqrt{6/\beta}/u_0$;

3) Approximation from $u_0$ only: Using \( u_0 \) only in gauge fixing while approximating \( \alpha_s \) via  $\alpha_s\simeq -\frac{4 \mathrm{ln} u_0}{3.0684}$~\cite{Alford:1995hw,Orginos:1998ue} which avoids to define $g_0$ from the action, and similar procedure gives:  $g^{(c)}_0=\sqrt{-\frac{16\pi \mathrm{ln} u_0}{3.0684}}u_0$.

Different definitions can differ at ${\cal O}(\alpha_s^2)$ according to perturbative lattice QCD~\cite{Lepage:1992xa}. Because tadpole improvement is essential for ensuring good convergence of the lattice perturbative series, the naive definition—though seemingly natural—is generally not suitable. For the MILC ensemble a06m310 at $a$=0.0566 fm with $m_{\pi}$=310 MeV and $\hat{\beta}=5/3\beta=6.72$, three definitions yield: $g_0^{(a)} = 1.2199,\ g_0^{(b)} = 1.3768,\ g_0^{(c)} = 1.2476$, respectively. In practice, we generate the random distribution $\tilde{P}(\tilde{\Lambda}^a(x))=\frac{1}{\sqrt{2\pi \tilde{\xi}}}\exp\left\{-\frac{1}{2\tilde{\xi}}[\tilde{\Lambda}^a(x)]^2\right\}$ for the dimensionless quantity \( \tilde{\Lambda}\equiv g_0 \Lambda a^2 \), meaning different \( g_0 \) definitions correspond to different effective gauge-fixing parameters \( \xi=\frac{1}{g_0^2}\tilde{\xi} \). 

In this work, we use $g_0^{(a)}$ to define the $\tilde{\xi}$ needed by $\xi=0,0.2,0.4,0.8,1.0$, and the effective $\xi$ with the other $g_0$ definition can be obtained with the rescale factor $(g^a_0/g_0)^2$.

The gauge-fixing algorithm we employ is the “over-relaxation” method described in Refs.~\cite{Mandula:1990vs,Giusti:2001xf,Schrock:2012fj,Bicudo:2015rma}, with its main procedure described as follows:\\

1) Separate all sites into two even-odd subsets, and start from the unitary gauge transformation matrix $G$ = 1.

2) Calculate the gauge fixing criterion
\begin{align}  
\theta = \frac{1}{N_c V} \sum_x \text{Tr}\left[ \Delta(x) \Delta^\dagger(x) \right],  
\end{align} 
where $V$ is the lattice volume.

3) For all the sites in the first parity subset:

3.1) Compute 
\begin{align} \label{eq:K}
    K[x;G]=&\sum_{\mu=1}^4 \left[U_\mu(x+\hat{\mu}\frac{a}{2})+U_\mu(x-\hat{\mu}\frac{a}{2})^{\dagger}\right]-{\rm i}\tilde{\Lambda}(x);
\end{align}

3.2) Decomposite $K$ into 3 $SU(2)$ subgroups $K_{\rm sub}$, construct 
\begin{align}
    \tilde{G}(x)=&\left.\left(A_0 \cdot I-\sum_{i=1}^3 A_i \cdot \sigma_i\right)\right|_{\text {renormalize}}\nonumber \\
    =&K_{\rm sub}(x)^{\dagger} / \sqrt{\operatorname{det} (K_{\rm sub}(x)^{\dagger})}
\end{align}
with 4 parameters $A_\mu$ extracted from $K_{\rm sub}$. One may apply over-relaxation by updating the parameters as  
 \( A_0' = \cos(\omega \cdot \arccos A_0) \), \( A_i' = A_i \frac{\sin(\omega \cdot \arccos A_0)}{\sin(\arccos A_0)} \).  While the over-relaxation parameter \(\omega\) is normally chosen between 1 and 2, we found that in the \(\xi\)-gauge, values of \(\omega > 1\) lead to poorer convergence in \(\theta\). For this reason, we fix \(\omega = 1\) in our simulations, which corresponds to performing no over-relaxation.Then update the gauge link and rotation
\begin{align}
  U_\mu(x+\hat{\mu}a/2) &\to \tilde{G}(x) \, U_\mu(x+\hat{\mu}a/2) \, \tilde{G}(x+\hat{\mu}a)^{\dagger},\nonumber\\
  G(x) &\to \tilde{G}(x) G(x);
\end{align}

3.3) Repeat the update for the other two SU(2) subgroups.

4) Repeat the step 3 for the other parity subset.

5) Repeat the step 2-4 until $\theta$ is smaller than the target precision.  

There is another widely used gauge fixing precision criteria for the Landau gauge by requiring
\begin{align}
    \delta(n) = F[G(n)] - F[G(n-1)]
\end{align}
to be smaller than given $\delta$, where 
\begin{align} \label{eq:F}
    F[G]=&- \sum_x\mathrm{Re}\mathrm{Tr}\big\{\sum_{\mu=1}^4 \left[G(x) U_\mu(x+\hat{\mu}\frac{a}{2}) G(x+\hat{\mu}a)^{\dagger} \nonumber \right. \\
    &\left.+G(x) U_\mu(x-\hat{\mu}\frac{a}{2})^{\dagger} G(x-\hat{\mu}a)^{\dagger}\right],
\end{align}
$G(n)$ represents the gauge rotation at the $n$-th step.  The gauge fixing  will stop at the $m$-th step once $\delta^F(m)$ is smaller than the preassigned value $\delta^F$, and the previous study on the MILC ensembles in the lattice spacing range $a\in[0.03,0.12]$ fm suggests that $\theta\sim 24 \delta^F\,$ in all the cases~\cite{Zhang:2024omt}. Landau gauge fixing in this work was performed with precisions \(\delta^F = 10^{-n}\) for integer \(n\), and then correspond to \(\theta \sim 2.4 \times 10^{-n+1}\).

To verify the implementation of the $\xi$-gauge fixing on the lattice, one must compare lattice-computed observables with known continuum results. A suitable class of observables are the $\xi$-dependent, off-shell quark matrix elements for bilinear operators, $\mathcal{O}_{\Gamma} \equiv \bar{\psi}\Gamma\psi$, specifically $\langle q|\mathcal{O}_{\Gamma}|q\rangle$.

In the continuum, the vector and axial-vector currents are protected from renormalization in the continuum as a consequence of current conservation and partially conserved axial current (PCAC) relations, respectively. Furthermore, chiral symmetry ensures that the renormalization of the pseudo-scalar current is identical to that of the scalar current. For a Euclidean momentum $p^2$, the scalar ($S\equiv \mathcal{O}_{\cal I}$) and tensor ($T_{\mu\nu}\equiv \mathcal{O}_{\sigma_{\mu\nu}}$) matrix elements are known to 3-loop order in dimensional regularization~\cite{Gracey:2003yr}:
\begin{align}
    \langle q^R(p)|\mathrm{S}|q^R(p)\rangle&=1+\frac{\alpha_sC_F}{4\pi}[3(\frac{1}{\bar{\epsilon}}+\mathrm{ln}\frac{\mu^2}{p^2})+4+\xi]\nonumber\\
    \quad\quad+{\cal O}(\alpha_s^2),\nonumber\\
     \langle q^R(p)|\mathrm{T}_{\mu\nu}|q^R(p)\rangle&=\sigma_{\mu\nu}\big\{1+\frac{\alpha_sC_F}{4\pi}[-(\frac{1}{\bar{\epsilon}}+\mathrm{ln}\frac{\mu^2}{p^2})-\xi]\nonumber\\
     &\quad\quad +{\cal O}(\alpha_s^2)\big\}.
\end{align}
Note that while the operator itself is gauge-invariant, the off-shell quark state used in its evaluation is defined in a specific gauge.

The logarithmic term $\frac{1}{\bar{\epsilon}}+\ln(\mu^2/p^2)$ is gauge-independent. On the lattice, this term is replaced by $-\ln(a^2 p^2) + c_{\cal O}$, where the constant $c_{\cal O}$ depends on the specific operator and the discretization of the fermion and gauge actions. In contrast, the finite $\mathcal{O}(\alpha_s)$ coefficient of $\xi$ is non-zero and operator-dependent due to the off-shell external state. Notably, for the scalar and tensor operators, this coefficient has opposite signs. Crucially, this specific gauge dependence is a property of the currents themselves in the off-shell quark state; it is therefore independent of the ultraviolet regularization and must match between the lattice and continuum formulations.

This fundamental feature is leveraged in the regularization-independent momentum-subtraction (RI/MOM) scheme~\cite{Martinelli:1994ty}, where a condition such as $Z^{\mathrm{RI}}_{{\cal O}_{\Gamma}} \langle q^R(p)|{\cal O}_{\Gamma}|q^R(p)\rangle \equiv \Gamma$ is imposed at a scale $p^2=\mu_0^2$. The subsequent perturbative matching from the gauge-dependent RI/MOM scheme to the gauge-independent $\overline{\mathrm{MS}}$ scheme is designed to cancel this $\xi$-dependence. However, a direct verification that the final $\overline{\mathrm{MS}}$ renormalization constants are indeed gauge-independent has, until now, been absent.

\begin{figure}[!h]
    \centering
    \includegraphics[width=0.95\linewidth]{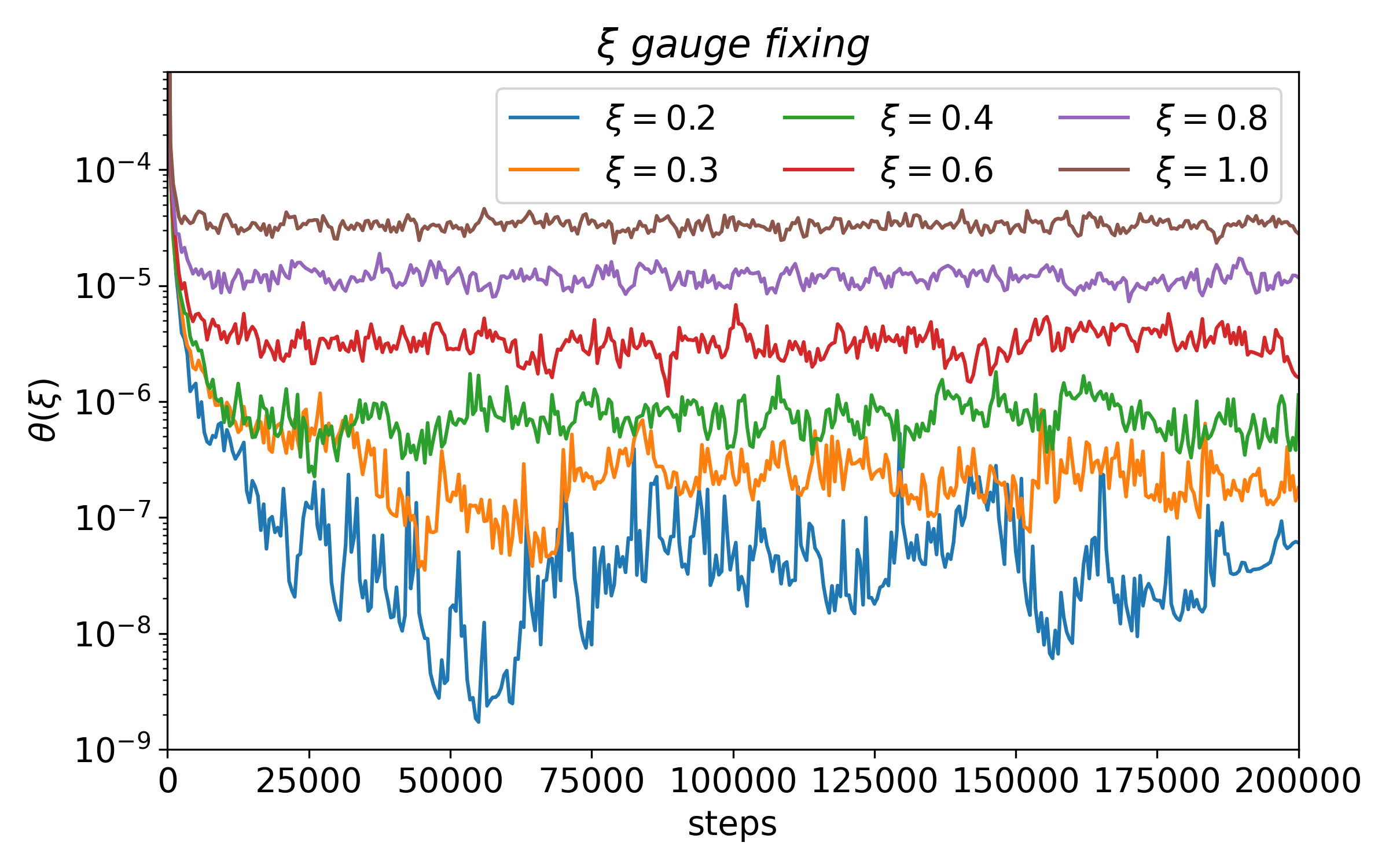}
    \includegraphics[width=0.90\linewidth]{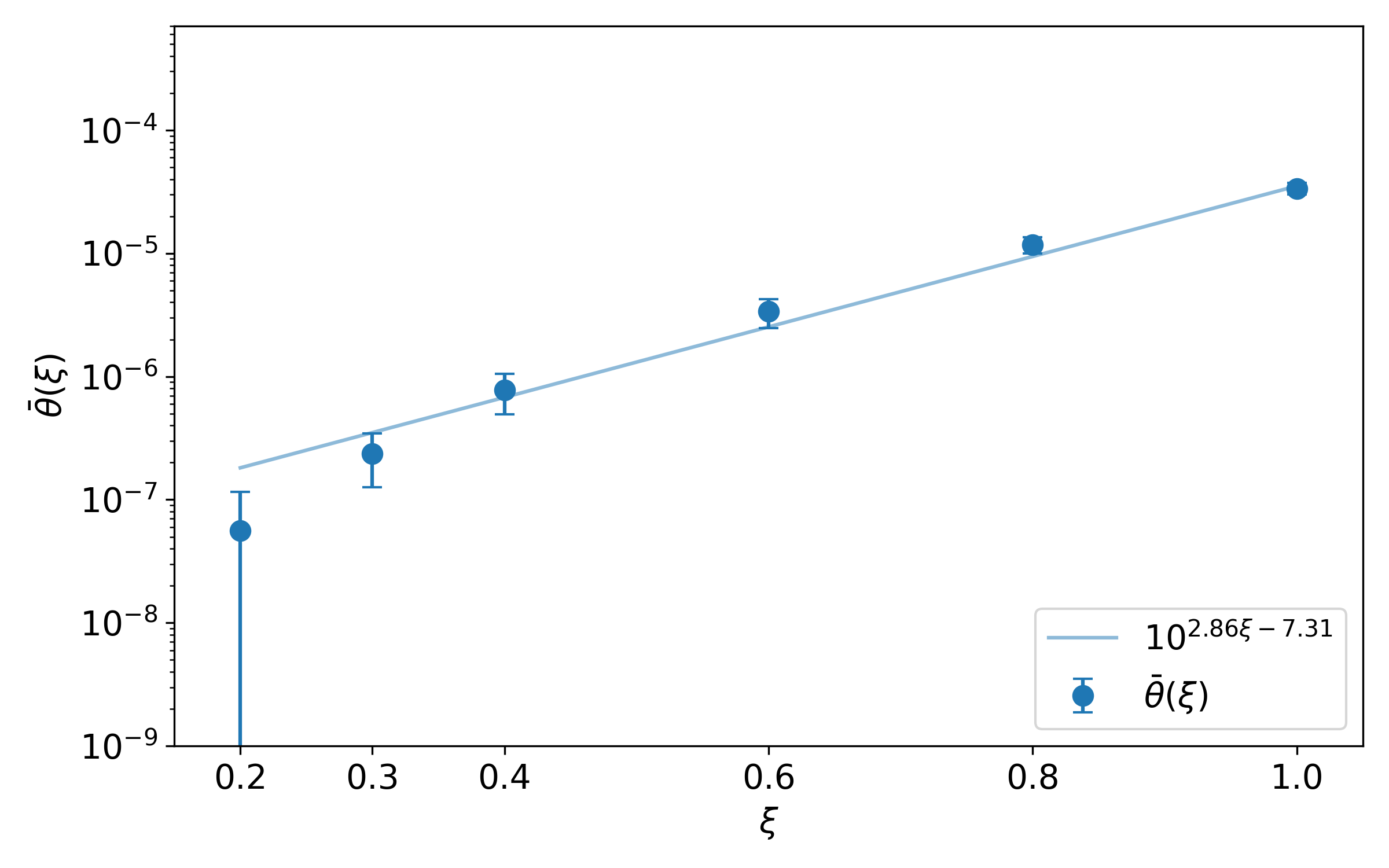}
    \caption{Upper panel: gauge fixing criteria $\theta$ as a function of iteration steps, for different $\xi$ using $g_0^{(a)}$ on a typical configuration of the a06m310 ensemble. Lower panel: Exponential growth of the gauge-fixing residual \(\bar{\theta}\) with \(\xi\), based on over the iteration window 100,001–200,000. (Please change $\theta^G_{\xi}$ into $\bar{\theta}(\xi)$)}
    \label{fig:fixing}
\end{figure}

\subsection{Numerical challenge in the $\xi$-gauge fixing}

As demonstrated in the upper panel of Fig.~\ref{fig:fixing} using the MILC ensemble a06m310 at the $a=0.06$ fm with the information detailed in the next section, The primary reason for this absence is that convergence to a small \(\theta(\xi)\) becomes unattainable at large \(\xi\), even after 100,000 iterations. The mean residual \(\bar{\theta}\), measured between steps 100,000–200,000, exhibits an exponential scaling with \(\xi\) (\(\bar{\theta}(\xi) \approx 10^{\,2.9\xi - 7.3}\)), as illustrated in the lower panel.
Nevertheless, the Landau gauge with $\xi=0$ is free of this convergence problem since $\Lambda=0$, and one can reach much higher precision likes $10^{-15}$. 

To ensures consistency across different precision levels, we generated a random \(\Lambda\) field (\(\Lambda\)=0 in the Landau gauge) for each gauge configuration and recorded the gauge-fixed links dynamically as the minimization process crossed each target accuracy \(\theta\). This approach is applied throughout all precision-dependence and extrapolation studies in this work.

\begin{figure}[!h]
    \centering
    \includegraphics[width=0.95\linewidth]{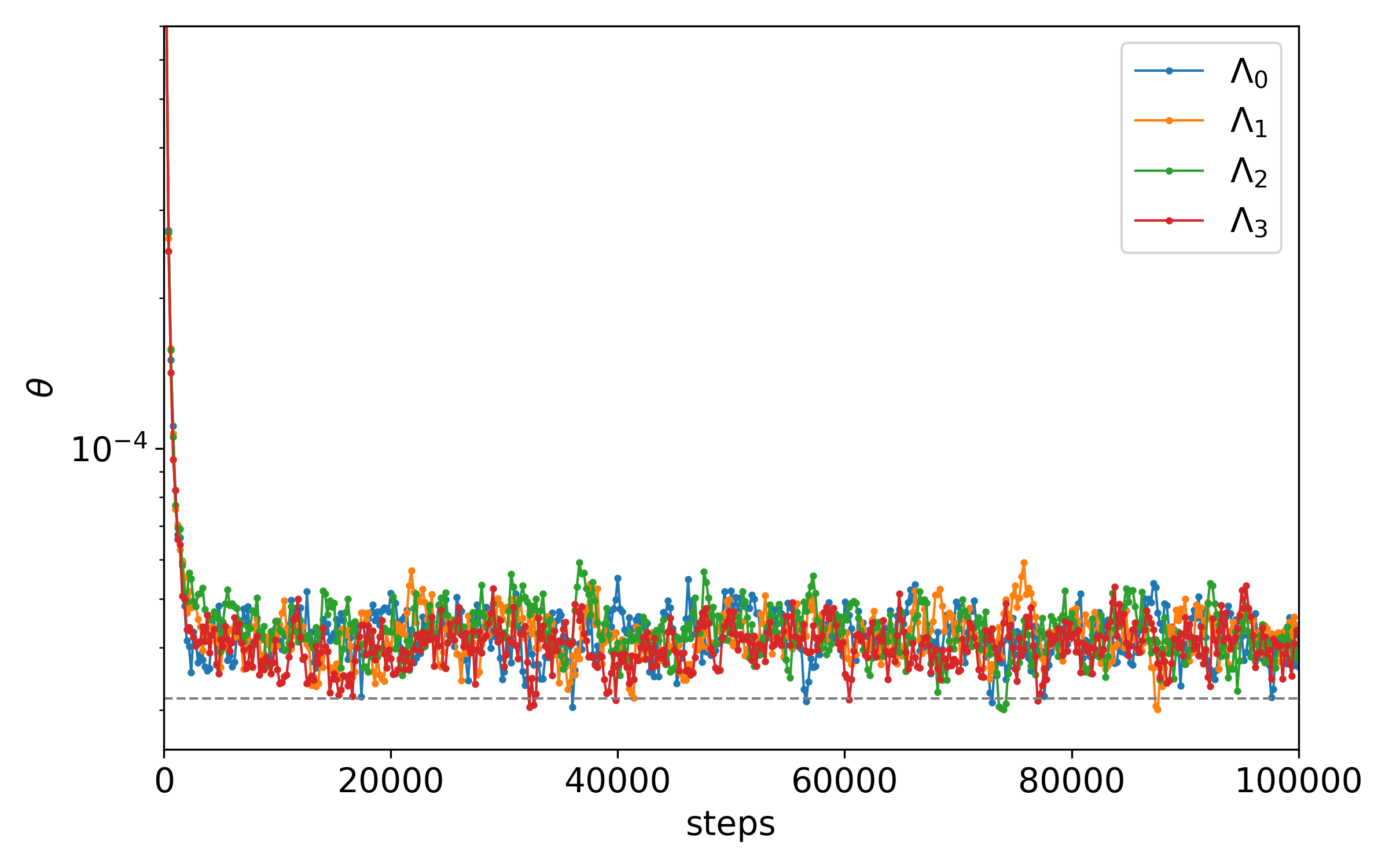}
    \includegraphics[width=0.95\linewidth]{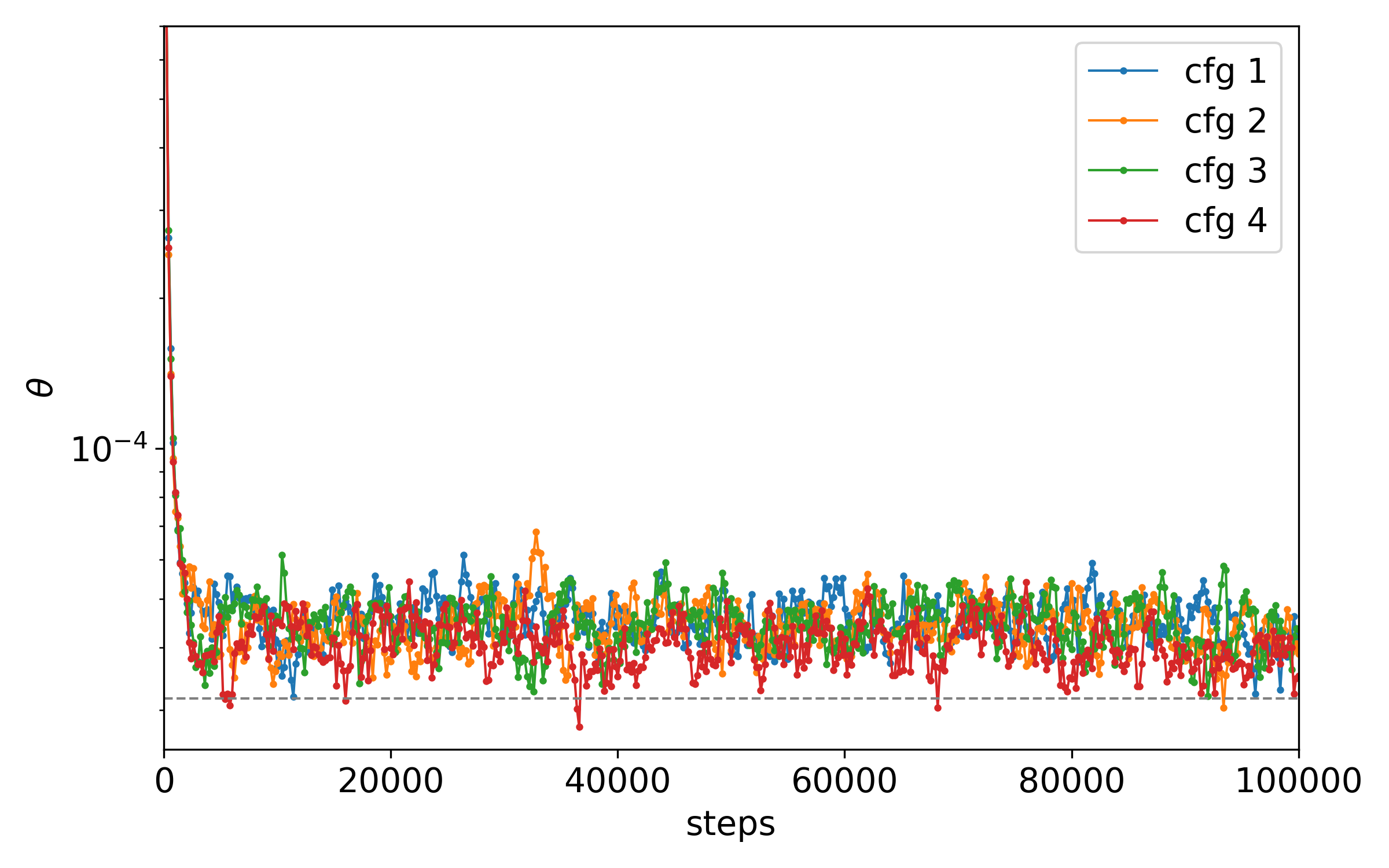}
    \caption{Convergence trajectories of the residual $\theta$ with $\xi=1$ for different $\Lambda$ on given configuration (upper panel) and different configurations with respective random $\Lambda$ (lower panel).}
    \label{fig:gfix_comp}
\end{figure}

We further verify that the convergence behavior is insensitive to both the choice of \(\Lambda\) and the specific gauge configuration:

1. Robustness against \(\Lambda\) choice: For a fixed gauge configuration, we tested multiple random initializations of the \(\Lambda\) field. As shown in the upper panel of Fig.~\ref{fig:gfix_comp}, all sampled \(\Lambda\) configurations for \(\xi = 1\) exhibit similar convergence trajectories. The fluctuations of $\theta$ ensure that the residual $10^{-4.5}$ (the lowest one used in the work, dash line in Fig.~\ref{fig:gfix_comp}) can eventually be reached for all sampled \(\Lambda\) , even though it lies slightly below the ``plateau” of \(\theta\). This indicates that the attainable precision is unaffected by the choice of \(\Lambda\).

2. Ensemble-wide consistency: As illustrated in the lower panel of Fig.~\ref{fig:gfix_comp}, we repeated the same test for different gauge configurations, and the conclusion remains unchanged.

Consequently, the resulting distribution of \(\Lambda\) functions should remain purely Gaussian, as no selection bias was introduced into the analysis.

In principle, \(\theta\) cannot be made arbitrarily small for sufficiently large \(\xi\) due to the gauge-fixing condition in Eq.~(\ref{eq:condition}). If we define $\tilde{\Delta}(x)\equiv g_0\Delta(x)$ and separate it as,
\begin{align}
    \tilde{\Delta}(x)&=\tilde{\Delta}_U(x)-\tilde{\Lambda}(x),\ \nonumber\\
    \tilde{\Delta}_U(x)&\equiv \sum_{\mu, \eta = \pm} \eta \left[ \frac{U_\mu(x + \eta \frac{\hat{\mu}}{2} a) - U_\mu^\dagger(x + \eta \frac{\hat{\mu}}{2} a)}{2 {\rm i} } \right]_{\text{Traceless}},\ \nonumber\\
    \tilde{\Lambda}(x)&\equiv g_0\Lambda a^2,
\end{align}
then $\tilde{\Delta}_U$ is bounded but $\tilde{\Lambda}\propto \sqrt{\xi}$ is not. Thus it is crucial to verify whether the ``plateau” of \(\theta\) originates from this concept issue.

\begin{figure}[!h]
    \centering
    \includegraphics[width=0.95\linewidth]{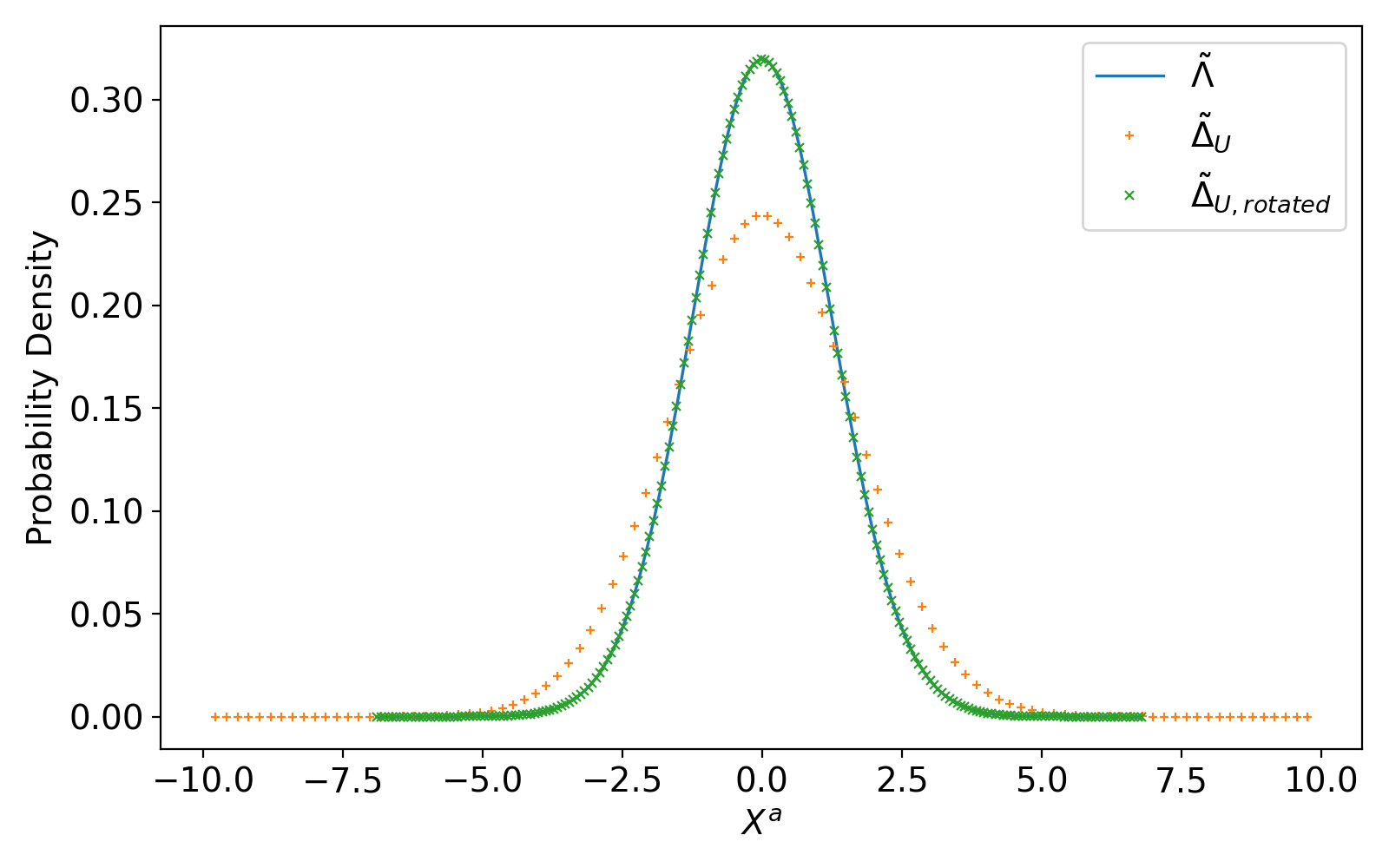}
    \caption{Probability density of $SU(3)$ components of the 
    $\tilde{\Delta}_U^a$ before (orange upright crosses) and after (green saltire crosses) gauge fixing, and that of $\tilde{\Lambda}^a$ (blue curve),
    in the case of $\xi=1.0$, $\theta=10^{-4.5}$. For each space-time points, 8 $SU(3)$ components $X^a$ are extracted through $X^a=2\ \mathrm{tr}{(Xt^a)}$.}
    \label{fig:p_ca}
\end{figure}

As shown in Fig.~\ref{fig:p_ca}, the distribution of $\tilde{\Delta}_U^a\equiv 2\mathrm{Tr}[t^a\tilde{\Delta}_U]$ without the gauge fixing after projected to the SU(3) generators can be rougly described by a gaussian distribution with variance $\sigma(\tilde{\Delta}^a_U)=1.63$. Since $\sigma(\tilde{\Delta}^a_U)$ is larger than that of $\tilde{\Lambda}^a\equiv 2\mathrm{Tr}[t^a\tilde{\Lambda}]$ with the largest $\xi=1.0$, $\sigma(\tilde{\Lambda}^a)=g_0\sqrt{\xi}=1.25$, the gauge fixing is a procedure to reduce the variance of $\tilde{\Delta}^a_U$ to match that of the $\tilde{\Lambda}^a$, likes the Landau gauge fixing. The practical calculation suggests that the largest $|\tilde{\Lambda}^a|$ which we generated in the $\xi=1.0$ case is around 7.2 which is around $5.8\sigma$ of its distribution, where that of $\tilde{\Delta}^a_U$ is around 8.8 which is larger. It means that practically we have not brought the theoretical concept  issue into real calculation.

\begin{figure}[!h]
    \centering
    \includegraphics[width=0.95\linewidth]{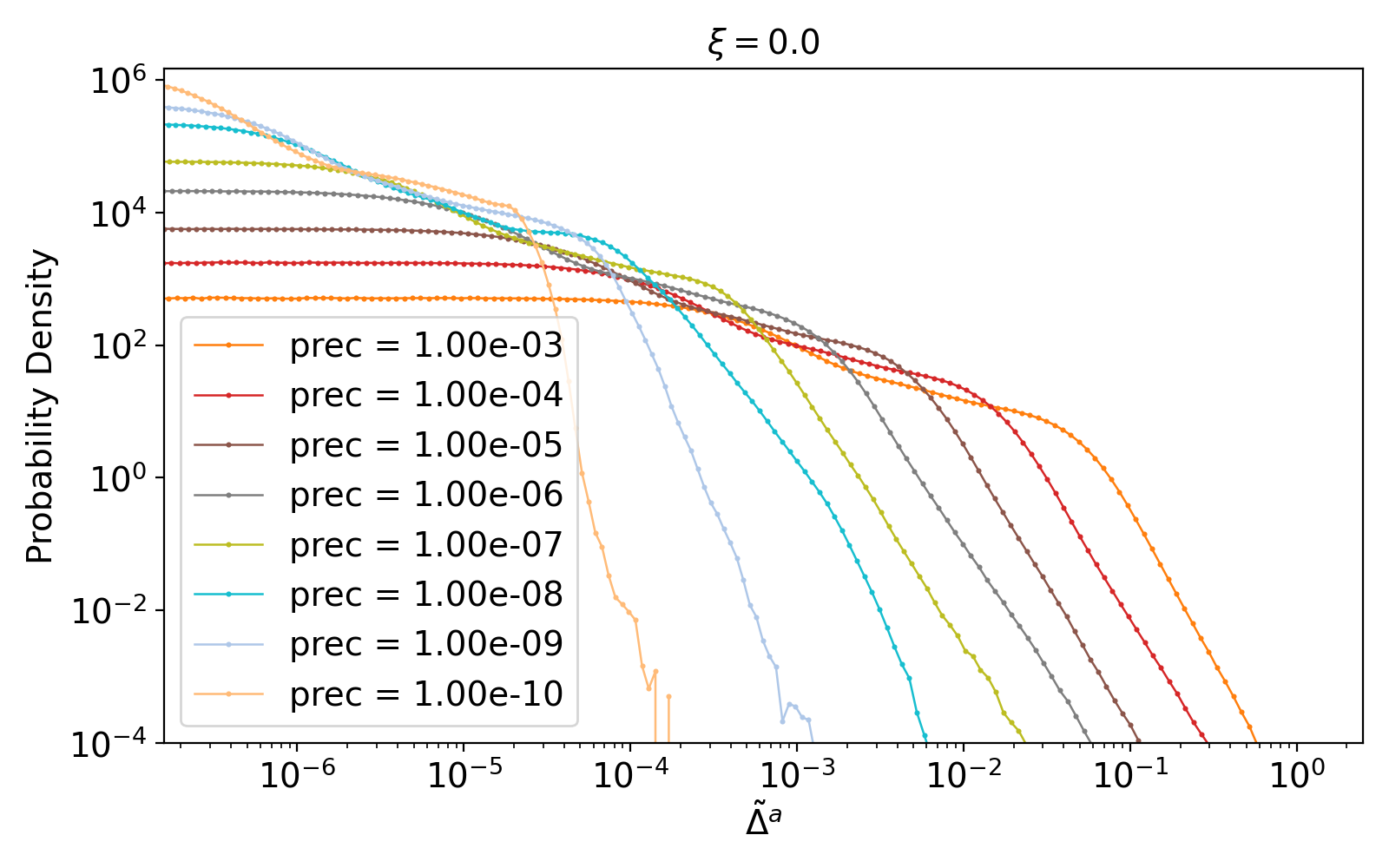}    
    \caption{Probability density of the residual $\tilde{\Delta}^a$ in Landau gauge.}
    \label{fig:residual_0}
\end{figure}

We further illustrate the distribution of the residual \(\tilde{\Delta}^a\) for different \(\theta\) values in the Landau gauge, shown in Fig.~\ref{fig:residual_0}. Taking the case \(\theta = 10^{-3}\) as an example, the distribution \(P(\tilde{\Delta}^a)\) is flat for \(\tilde{\Delta}^a \lesssim 10^{-3.5}\), decreases as \(\sim (\tilde{\Delta}^a)^{-1}\) for \(\tilde{\Delta}^a \lesssim 10^{-1}\), and then falls more rapidly as \(\sim (\tilde{\Delta}^a)^{-4}\) for larger \(\tilde{\Delta}^a\). As \(\theta\) decreases, the region of rapid suppression shifts toward smaller \(\tilde{\Delta}^a\), accompanied by a higher plateau value near \(\tilde{\Delta}^a \sim 0\). This behavior suggests that imprecise Landau gauge fixing affects physical quantities differently than the \(\xi\) gauge with small but non-zero \(\xi\), where deviations from the Landau gauge are expected to follow a Gaussian distribution which decays exponentially at relatively large $\tilde{\Delta}^a$.

\begin{figure}[!h]
    \centering
    \includegraphics[width=0.95\linewidth]{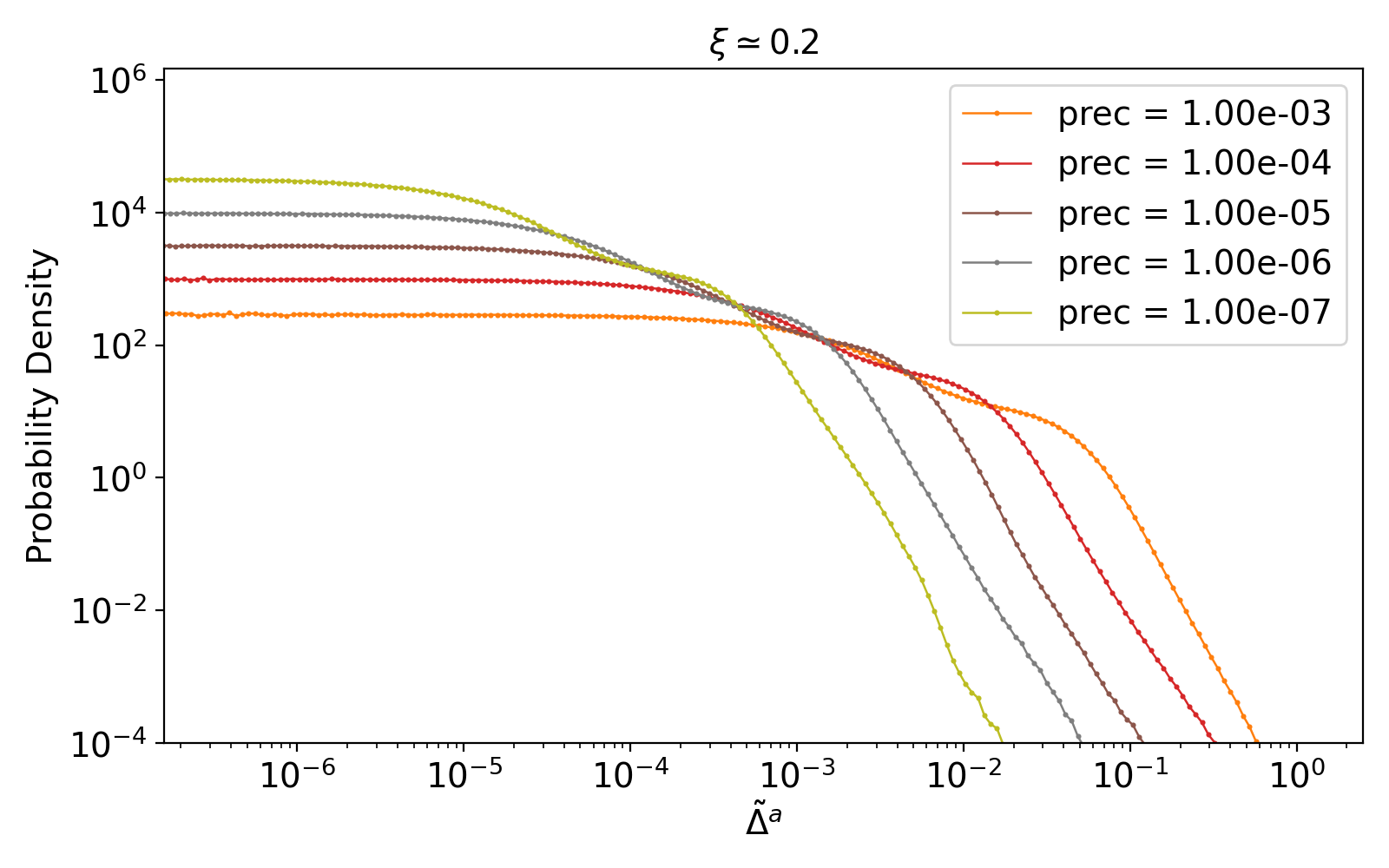}
    \includegraphics[width=0.95\linewidth]{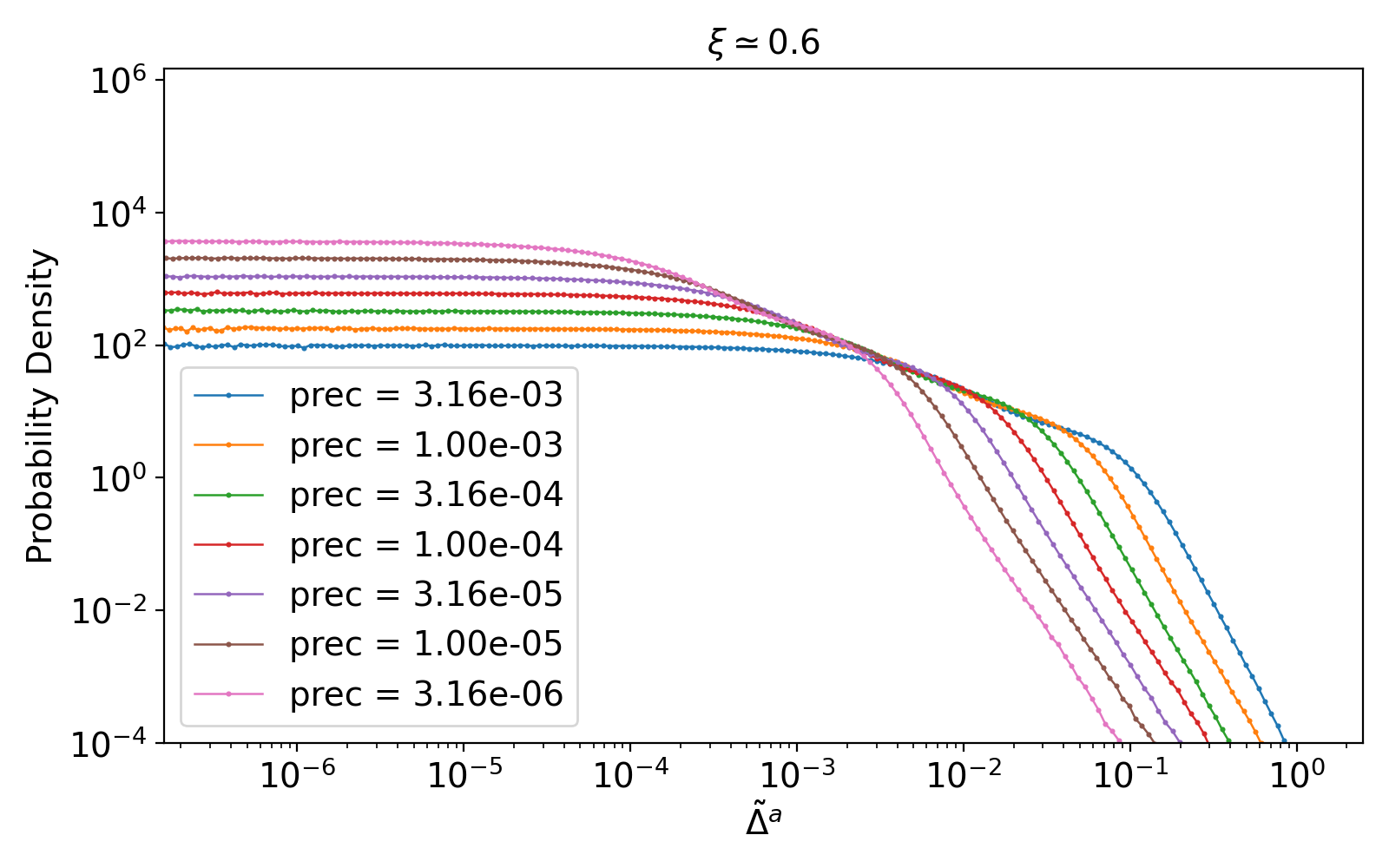}
    \includegraphics[width=0.95\linewidth]{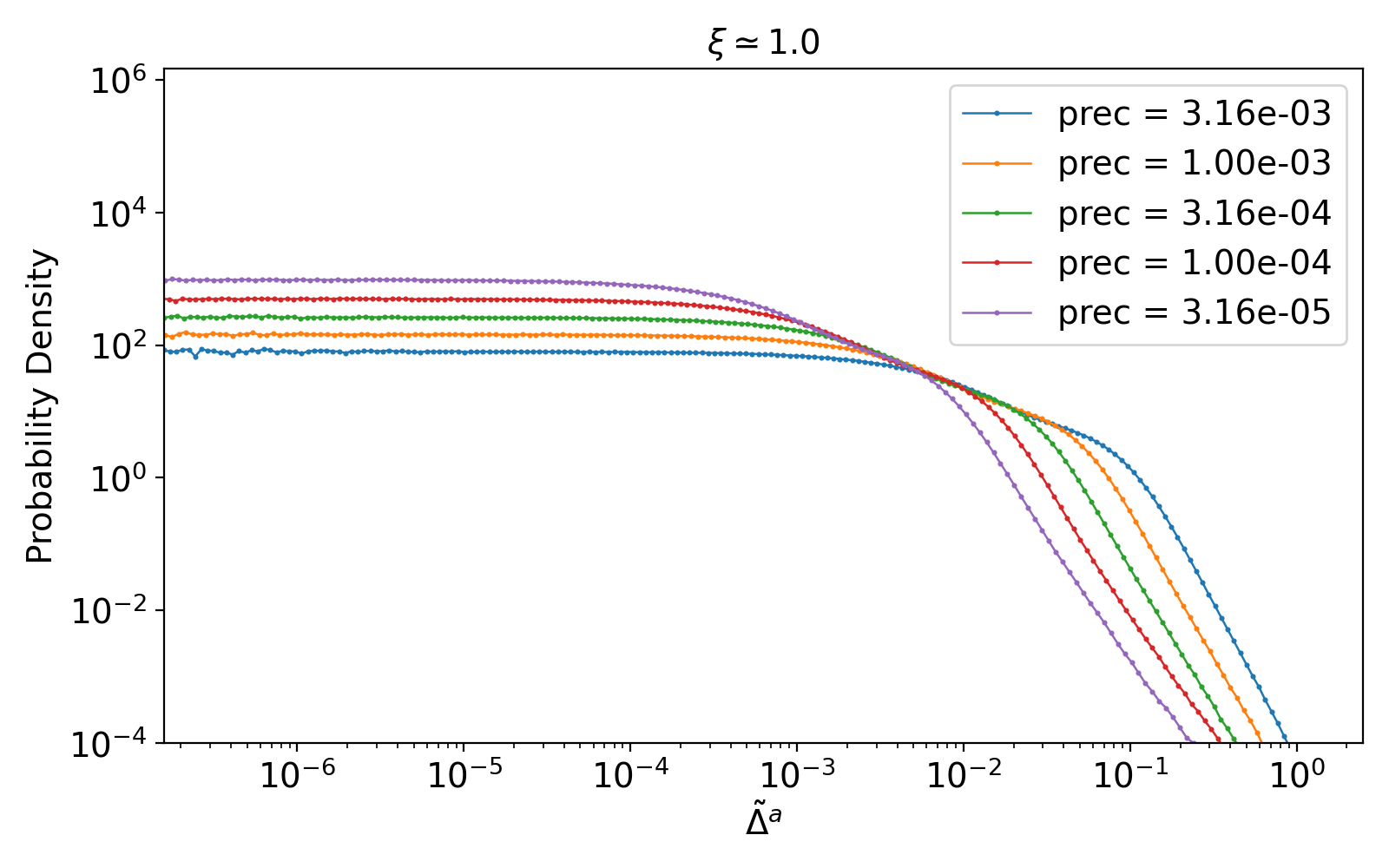}
    \caption{Probability density of the residual $\tilde{\Delta}^a$ with different precision (prec)), for $\xi\simeq0.2$, 0.6 and 1.0.}
    \label{fig:residual}
\end{figure}

Based on similar calculations for \(\xi \simeq 0.2\), \(0.6\), and \(1.0\), shown in Fig.~\ref{fig:residual}, the distributions of \(\tilde{\Delta}^a\) for non-zero \(\xi\) exhibit similar patterns in both their \(\tilde{\Delta}^a\) and \(\theta\) dependencies. This observation suggests that the empirical formula established for the Landau gauge may also describe the \(\theta\) dependence of physical quantities for \(\xi \le 1\), although we emphasize that this remains an assumption given the ill-posed nature of \(\xi\)-gauge fixing and the absence of a rigorous proof.

Ref.~\cite{Zhang:2024omt} demonstrated that the bias introduced by imprecise Landau gauge fixing on non-local operators can be captured by an empirical formula:
\begin{align}\label{eq:empi}
    X(\theta) = X^{\rm fit} \, e^{-c(X)\, \theta^{n(X)}},
\end{align}

where $X^{\rm fit},\ c$ and $n$ are fit parameters, and the fitting result of $X^{\rm fit}$ denotes the extrapolated exact result under perfect gauge fixing. For both the gauge link \(W(z) \equiv U_z(0,z)\) and the non-local quark bilinear \(\bar{\psi}(0)\gamma_t U_z(0,z)\psi(z)\), Ref.~\cite{Zhang:2024omt} founds \(n \sim 0.5\) and \(c \propto (z/a)^2\), with \(c\) reaching \(\mathcal{O}(30)\) at \(z \sim 1\ \text{fm}\) and \(a \sim 0.06\ \text{fm}\). The empirical form in Eq.~\eqref{eq:empi} remains valid up to \(\theta \sim 0.01\). 

\begin{figure}
    \centering
    \includegraphics[width=0.95\linewidth]{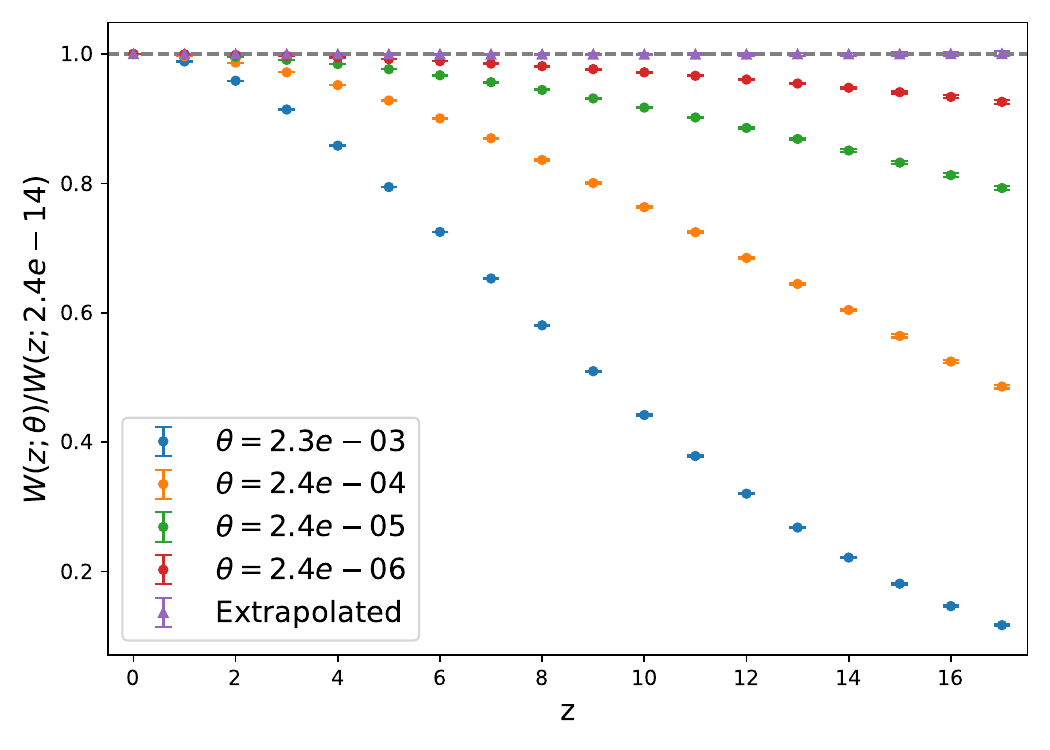}
    \caption{Ratio of the gauge link \(W(z) \equiv U_z(0,z)\) at various Landau gauge-fixing precisions \(\theta = \{2.3\times10^{-3}, 2.4\times10^{-4}, 2.4\times10^{-5}, 2.4\times10^{-6}\}\) to its high-precision value \(W(z; 2.4\times10^{-14})\). The ratio for the precision-extrapolated result (purple triangles, using the empirical formula) is also shown.}
    \label{fig:link}
\end{figure}

As shown in Fig.~\ref{fig:link}, fitting the parameters \(W(z;0)\), \(c_W\), and \(n_W\) using data at finite precisions \(\theta = \{2.3\times10^{-3}, 2.4\times10^{-4}, 2.4\times10^{-5}, 2.4\times10^{-6}\}\) yields a value of \(W(z;0)\) that agrees with the result obtained at the much higher precision \(\theta = 2.4\times10^{-14}\) within statistical uncertainties, despite the large deviations present in the data at the coarser precisions. 

Although this power-divergent bias vanishes for local operators (\(z = 0\)), the success of the precision extrapolation using only relatively poor gauge fixing data suggests that a similar strategy could be highly beneficial in the \(\xi\)-gauge, where achieving high-precision gauge fixing is currently impractical.

\section{Precision Extrapolation under Landau Gauge on Local Operators}\label{sec:precextralocal}

We first verify Eq.~(\ref{eq:empi}) and the precision extrapolation in the Landau gauge as a prerequisite for the $\xi$-gauge study. This verification uses the RI/MOM constants computed across a range of gauge-fixing precisions. In this study, we employ configurations generated by the MILC Collaboration~\cite{MILC:2010pul,MILC:2012znn,Bazavov:2017lyh}, utilizing the $2+1+1$ HISQ (Highly Improved Staggered Quark) fermion action and the one-loop Symanzik-improved gauge action. The specifics of these configurations are detailed in Tab.~\ref{tab:ensemble}. For the valence quarks, we use both clover and overlap fermion actions across these ensembles with the pion mass tuned to the same as that of light sea quark. Further details will be elaborated upon later in this Section.

\begin{table}
\begin{tabular}{c c c c c c}
\hline
\text{Action}  &\text{Symbol} & $6/g^2$ & $L^3 \times T$   & $a$ (fm) & $m_{\pi,{\rm ss}}$ (MeV)  \\
\hline   
HISQ+S$^{(1)}$ & a12m310      & 3.60    & $24^3\times\ 64$ & 0.1222   & 310                       \\
HISQ+S$^{(1)}$ & a09m310      & 3.78    & $32^3\times\ 96$ & 0.0879   & 310                       \\
HISQ+S$^{(1)}$ & a06m310      & 4.03    & $48^3\times 144$ & 0.0566   & 310                       \\
\hline
\end{tabular}
\caption{Information of the 2+1+1 flavor MILC ensembles~\cite{MILC:2010pul,MILC:2012znn,Bazavov:2017lyh} used in this study. The symbol S$^{(1)}$ denotes the Symanzik gauge action with full one-loop improvement, while the sea quark action employs the HISQ (Highly Improved Staggered Quark) discretization.}
\label{tab:ensemble}
\end{table}

\subsection{Renormalization Constants of $Z_{S,T}$ on Various Momentum}\label{subsec:rimom}

With point source quark propagators, one can define bare Green's function as:
\begin{align}
    G_{\mathcal{O}}\left(p_1, p_2\right)=\sum_{x, y} e^{-i\left(p_1 \cdot x-p_2 \cdot y\right)}\langle\psi(x) \mathcal{O}(0) \bar{\psi}(y)\rangle,
\end{align}
and then the amputated Green's function is generally defined as:
\begin{align}
    \Lambda_\mathcal{O}(p_1, p_2) = S^{-1}(p_1)G_\mathcal{O}(p_1, p_2)S^{-1}(p_2),
\end{align}
where $S^{-1}(p)$ represents the inverse of the quark propagator $S(p)\equiv \sum_{x} e^{-i p \cdot x}\langle\psi(x) \bar{\psi}(0)\rangle$ with momentum $p$. Following the LSZ reduction formalism, the RI/MOM renormalization constant is given by,
\begin{align}
     Z_{\mathcal{O}_{\Gamma}}(\mu)&\equiv \frac{Z_q(\mu)}{\frac{1}{12}\mathrm{Tr}[\Lambda_{\mathcal{O}}(p, p)\Gamma]}|_{\mu^2=p^2},\nonumber\\
     Z_q(\mu)&\equiv \frac{\mathrm{Tr}[p\!\!\!/ S^{-1}(p)]}{12p^2}|_{\mu^2=p^2}.
\end{align}
rather than directly computing $z_\mathcal{o}/z_q$, we evaluate $z_\mathcal{o}/z_v$ instead to circumvent the explicit use of $z_q$ which is subject to significant discretization errors, and extract $z_v$ from the vector current conservation condition of the pseudoscalar meson. further details on these operators are available in ref.~\cite{he:2022lse}. 

\begin{figure}
    \centering
    \includegraphics[width=0.93\linewidth]{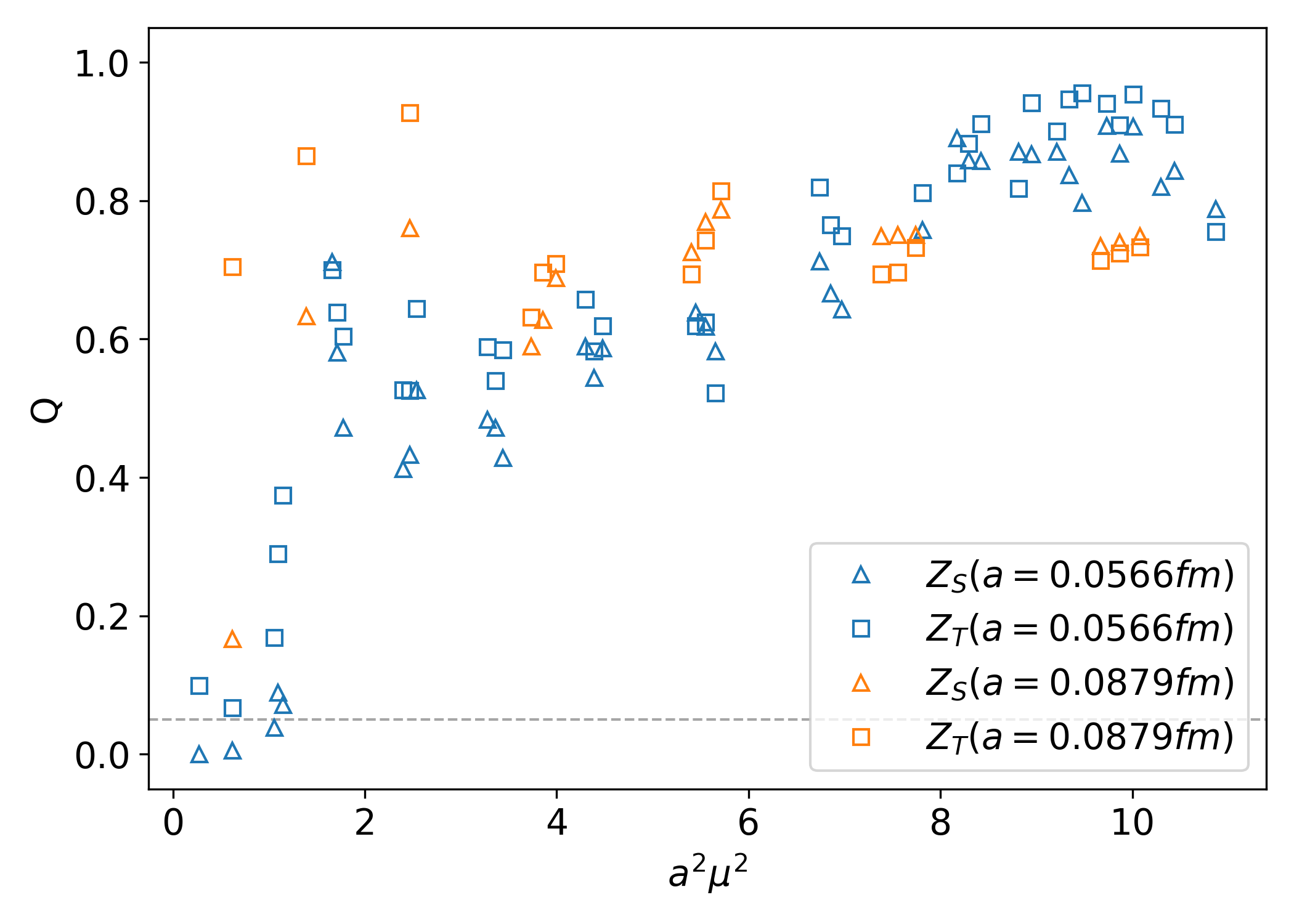}
    \includegraphics[width=0.95\linewidth]{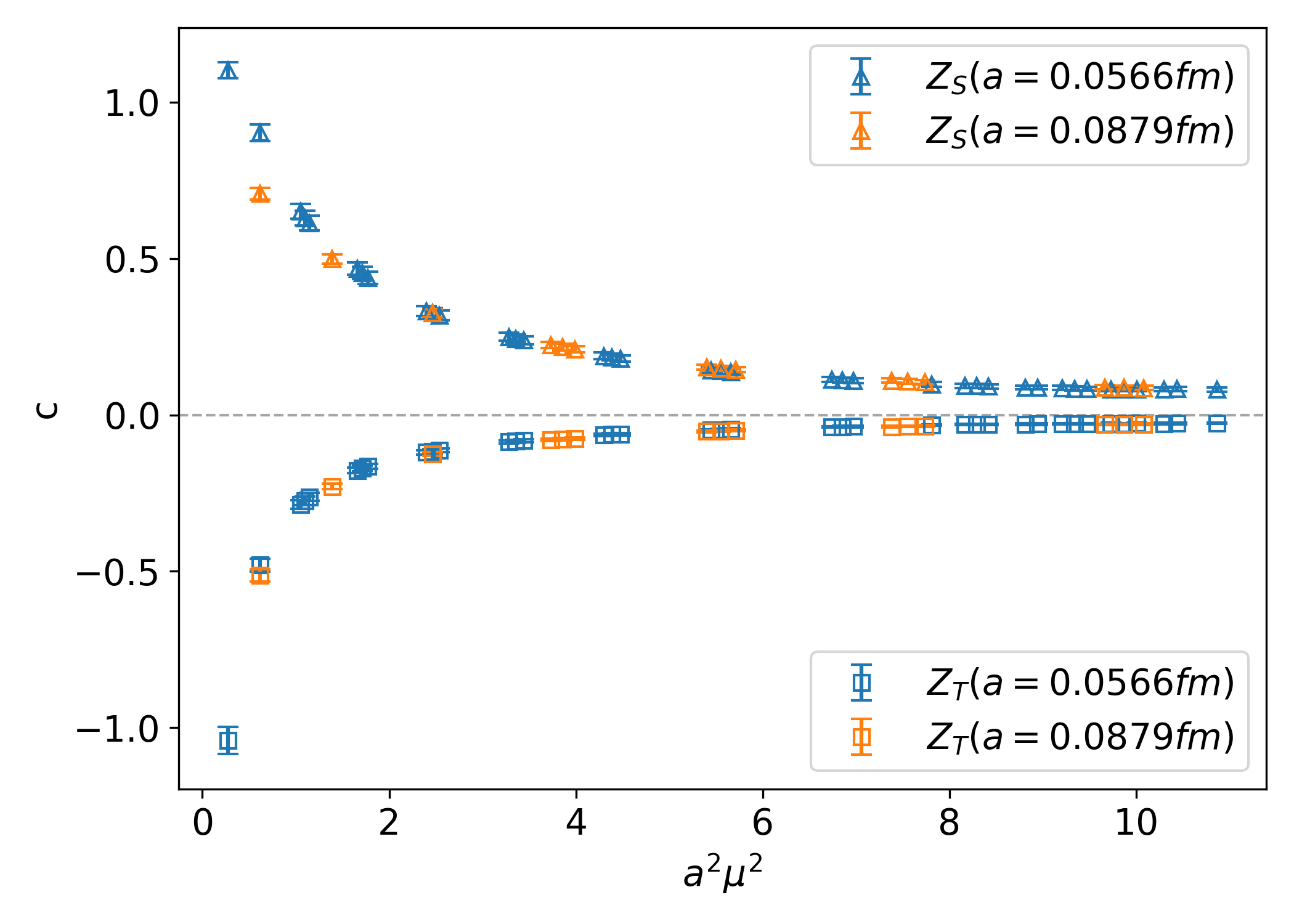}
    \includegraphics[width=0.95\linewidth]{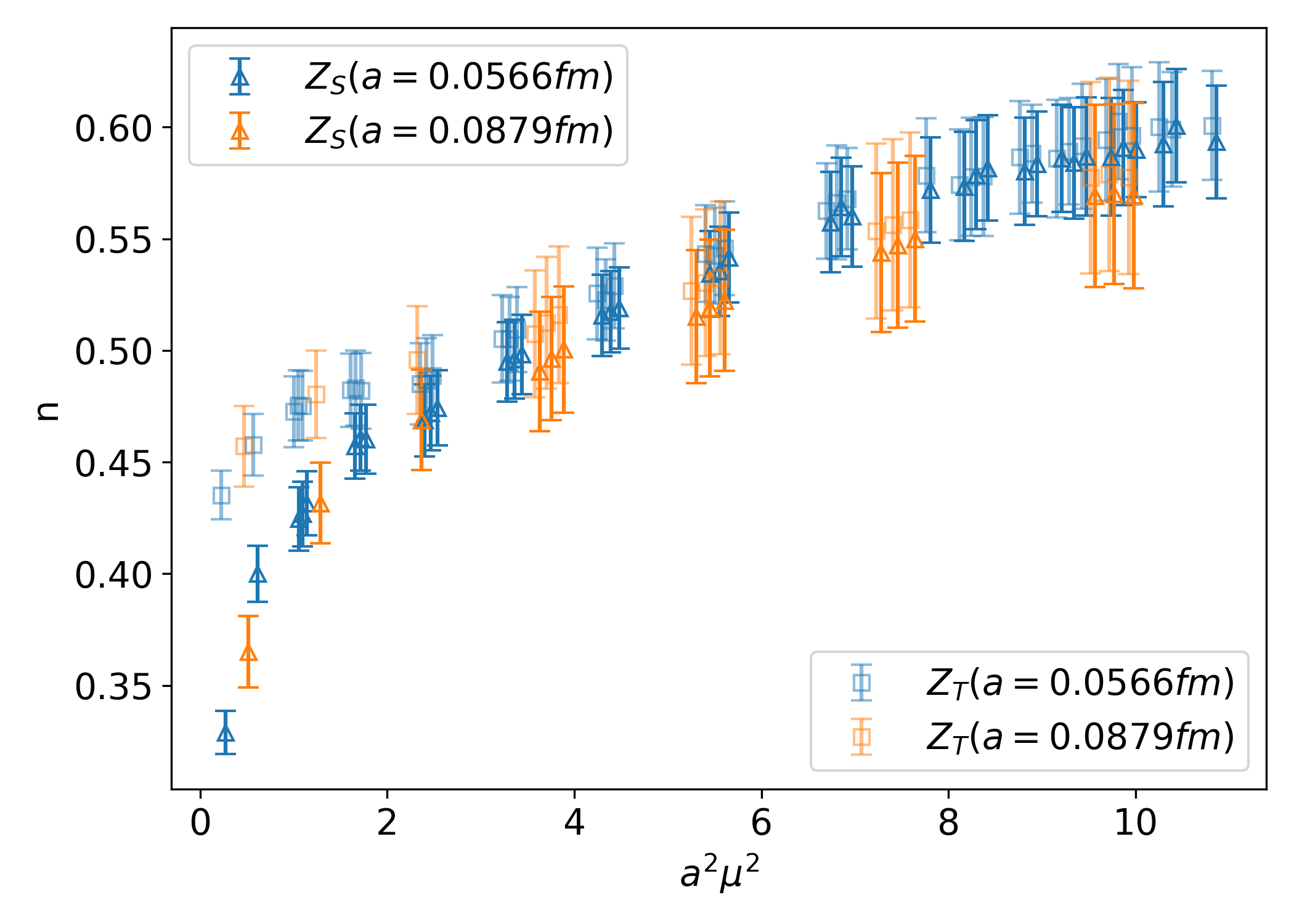}
    \caption{Q-values (upper panel), fit parameters $c$ (medium panel) and $n$ (lower panel) for the ratio $X(\theta)/X^{\rm fit}=e^{-c\cdot \theta^n}$ of $Z_S/Z_V$ and $Z_T/Z_V$ across various momenta $p$ and lattice spacing $a$. Original data is computed with point source propagators rotated by gauge rotation matrices corresponding to gauge fixing precisions $\theta = \{2.2\times10^{-1}, 2.3\times10^{-3}, 2.4\times10^{-4}, 2.4\times10^{-5}, 2.4\times10^{-6}, 2.4\times10^{-7}, 2.5\times10^{-9}, 2.4\times10^{-11}, 2.4\times10^{-14}\}$. For a given operator (scalar or tensor), both $c$ and $n$ lie on the same $a^2p^2$ curve, and insensitive to the lattice spacing $a$.  
    }
    \label{fig:zozv}
\end{figure}

For this analysis, we utilize valence overlap fermions~\cite{Chiu:1998gp} on two ensembles, a09m310 ($a\sim$ 0.09 fm) and a06m310 ($a\sim$ 0.06 fm). To investigate the momentum dependence, we employ point-source propagators and compute $Z_S/Z_V$ and $Z_T/Z_V$ at different $a^2p^2$. The propagators are solved using the deflated CG algorithm~\cite{Li:2010pw} with a residual tolerance of $1 \times 10^{-7}$, a precision demonstrated to be sufficient in previous renormalization studies. These quantities are derived by applying gauge rotations of varying precision $\theta \in \left(2.4\times10^{-14},2.2\times10^{-1}\right)$ to the same underlying quark propagators, and fitted with Eq.~(\ref{eq:empi}) to verify this fit ansatz. The Q values shown in the upper panel of Fig.~\ref{fig:zozv} suggest that fit qualities are acceptable except the lowest two momenta of $Z_S/Z_V$ at $a=0.0566$ fm. The correlation between the data with different theta has been taken into account through the correlation matrix in the precision extrapolation. 

The results of the fitted $c$ and $n$ are presented in the other two panels of Fig.~\ref{fig:zozv}. For a given operator (scalar or tensor), both $c$ and $n$ lie on the same $a^2p^2$ curve, and insensitive to the lattice spacing $a$. The approximation $n \approx 0.5$~\cite{Zhang:2024omt} seems to be held in large range of momenta, while $n$ decreases to roughly 0.4 at lower momenta. At the same time, the absolute value of $c$ increases rapidly at small $a^2p^2$. Both features require higher gauge-fixing precision at small momenta for a given relative deviation, and a smaller lattice spacing further tightens this requirement at fixed momentum. For instance, with gauge-fixing precision $\theta \sim 2 \times 10^{-6}$, $Z_S$ at $\mu = 1.75\,\textrm{GeV}$ deviates by 1.5\% for $a = 0.0566\,\textrm{fm}$ ($a^2\mu^2 \simeq 0.25$), whereas the deviation drops to 0.6\% for $a = 0.0879\,\textrm{fm}$ at the same scale and precision.

\begin{figure}
    \centering
    \includegraphics[width=0.95\linewidth]{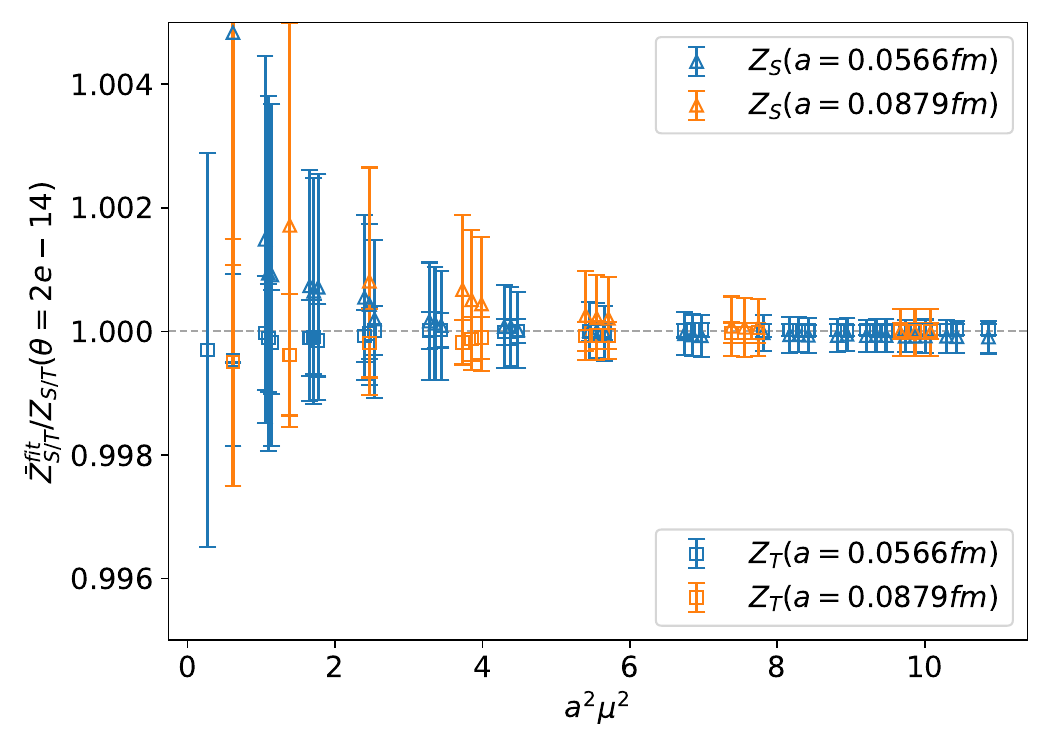}
    \caption{Ratio of precision-extrapolated $\bar{Z}^{fit}_{S,T}$ at various $a^2p^2$ to its high-precision value \(Z_{S,T}(2.4\times10^{-14})\). Only the results with relatively poor gauge fixing precision $\theta = \{2.2\times10^{-1}, 2.3\times10^{-3}, 2.4\times10^{-4}, 2.4\times10^{-5}\}$ are used for the extrapolation.}
    \label{fig:extra_z}
\end{figure}

The precision extrapolation allows us to correct for these deviations using data obtained even at relatively coarse gauge-fixing precisions. For the scalar and tensor operators, we fit the renormalization constants \(Z(a^2p^2,\theta)\) using the four largest values of \(\theta\) (\(\theta = 2.2\times10^{-1}, 2.3\times10^{-3}, 2.4\times10^{-4}, 2.4\times10^{-5}\)) based on the ansatz in Eq.~(\ref{eq:empi}), with \(\bar{Z}^{\rm fit}(a^2p^2)\), \(\bar{c}(a^2p^2)\), and \(\bar{n}(a^2p^2)\) as the renamed free parameters.
. The quality of the extrapolation is assessed by comparing the extrapolated values \(\bar{Z}^{\rm fit}(a^2p^2)\) to the high-precision results computed at \(\theta = 2.4\times10^{-14}\). As shown in Fig.~\ref{fig:extra_z}, the ratio of these values is consistent with unity within 0.2\% or even smaller statistical uncertainties.

\begin{figure}[!h]
    \centering
        \includegraphics[width=0.90\linewidth]{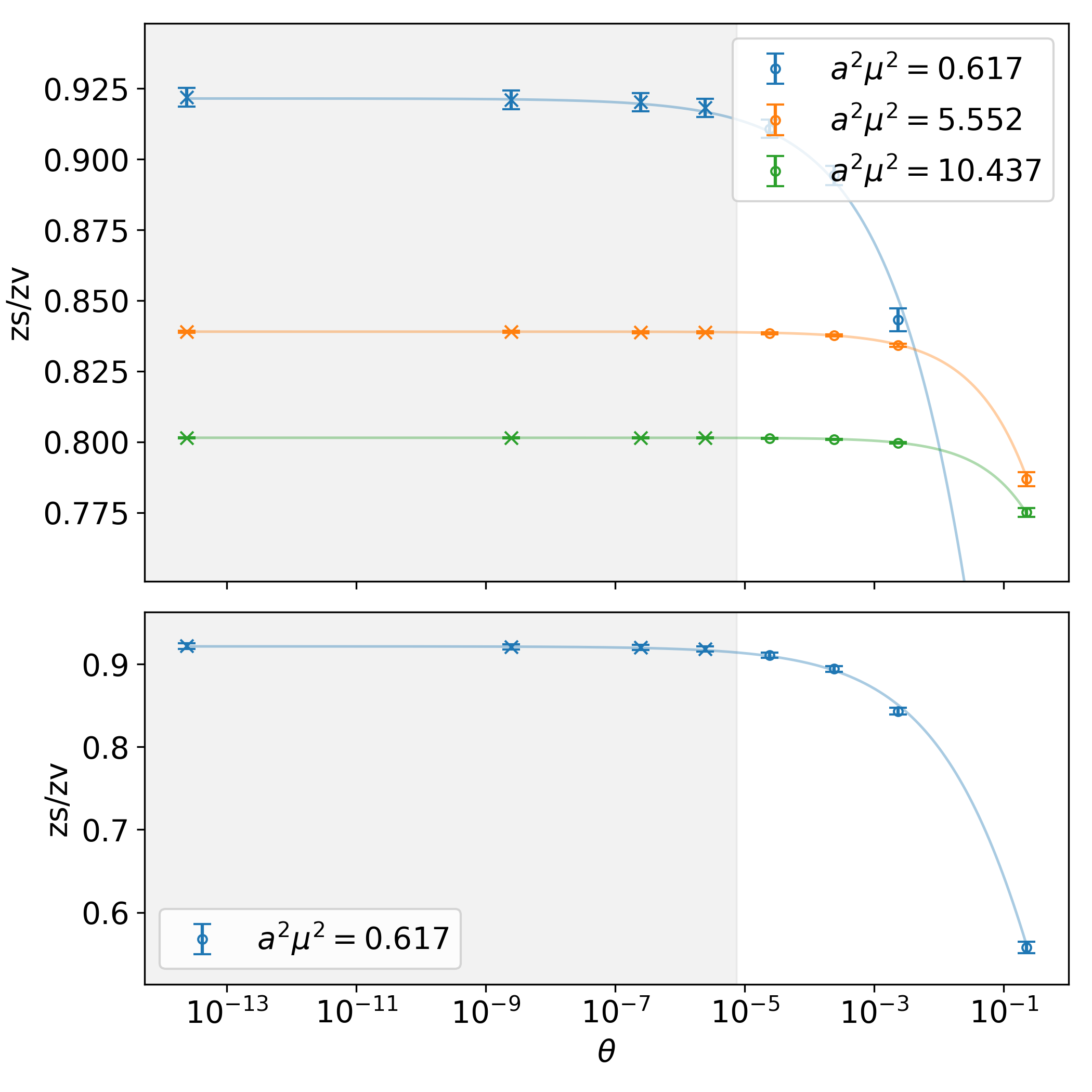}
        \includegraphics[width=0.90\linewidth]{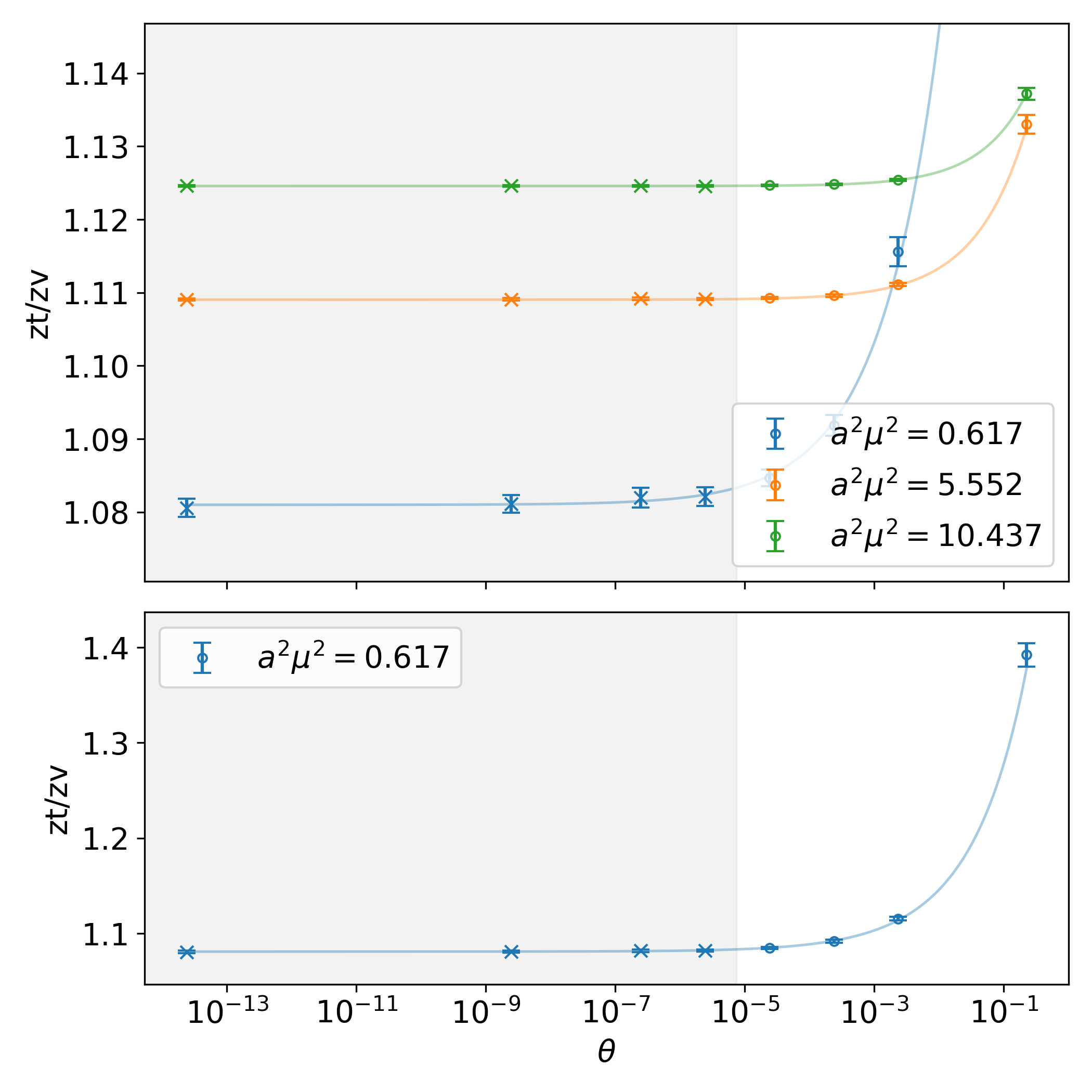}
    \caption{Gauge fixing residual $\theta$ dependencies of $Z_S/Z_V$ (upper panel) and $Z_T/Z_V$ (lower panel) for the scale $\mu$ in the infrared ($a^2\mu^2$=0.617), intermediate ($a^2\mu^2$=5.552), and ultraviolet ($a^2\mu^2$=10.437) regions. The gray-shaded region of \(\theta\) is excluded from the precision extrapolation, yet the data points within it agree well with the extrapolated curves (shown in color).}
    \label{fig:fit_check}
\end{figure}

In Fig.~\ref{fig:fit_check} (Fig. 9 of the revised manuscript), we explicitly show the data for \(Z(a^2p^2,\theta)\) as a function of \(\theta\) for three representative momenta spanning the infrared, intermediate, and ultraviolet regions. The data points marked with circles correspond to the data used in the extrapolation, while the solid curves represent the central values of the fits using Eq. (12). The data points indicated with crosses, which lie in the gray-shaded region excluded from the fit, are also shown and agree well with the extrapolated curves, further demonstrating that the fitting procedure captures the \(\theta\)-dependence reliably.

\subsection{Further Check with Valence Clover Volume Source Propagators}\label{subsec:momfrac}

For more accurate check on the deviation of imprecise gauge fixing for different operators, we generate volume source propagators with dimensionless momentum (5,5,0,0) (corresponds to $\mu\simeq$ 3 GeV) and gauge fixing precisions $\theta \in \left(2.4\times 10^{-14}, 2.5\times 10^{-2}\right)$, using valence clover fermions on two ensembles, a06m310 ($a\sim$ 0.06 fm and then $a^2p^2=0.85$) used above and also a12m310 ($a\sim$ 0.12 fm and then $a^2p^2=3.43$). The volume-source propagators are computed using the multigrid algorithm~\cite{Clark:2016rdz} to a similar tolerance of $\sim 1 \times 10^{-7}$, employing a setup consistent with Ref.~\cite{Zhang:2024omt}. Those propagators allows us to compute $Z_T/Z_V$ and also those of the quark energy moment tensor operators,
\begin{align}
    X_{2a}&\equiv\bar{\psi} \gamma_{\{\mu} \stackrel{\leftrightarrow}{D}_{\nu\}} \psi|_{\mu\neq \nu}, \nonumber \\
    X_{2b}&\equiv\frac{1}{2} \bar{\psi}\left(\gamma_1 \stackrel{\leftrightarrow}{D}_1+\gamma_2 \stackrel{\leftrightarrow}{D}_2-\gamma_3 \stackrel{\leftrightarrow}{D}_3-\gamma_4 \stackrel{\leftrightarrow}{D}_4\right) \psi
\end{align}
where the symmetric covariant derivative is given by $\stackrel{\leftrightarrow}{D}_{\nu} = \stackrel{\leftarrow}{D}_{\nu} - \stackrel{\rightarrow}{D}_{\nu}$. The use of volume source propagators enables the efficient calculation of $Z_{X_{2a(b)}}$, as all necessary derivative directions can be obtained at the sink of a single propagator. However, a full investigation of the $a^2p^2$ dependence requires repeated calculations at multiple momenta; such an analysis is beyond the scope of this work and is therefore omitted.

\begin{table}
\begin{tabular}{cc | c c c | c}
\cline{3-6}
  & & $c$ & $n$ &  $Z^{fit}$ & $Z(\epsilon_0)$ \\
\hline   
&$Z_T$&-0.80(4)&0.522(9)& 1.0916(2)&1.0915(1)\\
a06m310 &$Z_{X_{2a}}$&-1.60(5)&0.481(6)& 1.2376(4) & 1.2380(4)\\
&$Z_{X_{2b}}$&-1.64(5)&0.481(6)& 1.2249(4) & 1.2253(4)\\
\hline
&$Z_T$&-0.185(6)&0.513(9) & 1.0578(1)&1.0575(1)\\
a12m310 &$Z_{X_{2a}}$&-0.508(9)&0.482(5) & 1.1539(3) & 1.1542(2) \\
&$Z_{X_{2b}}$ &-0.467(9)&0.487(6) & 1.1338(3) & 1.1340(2)\\
\hline
\end{tabular}
\caption{Gauge fixing precision parameters $c$, $n$ and $Z^{fit}$ for $X(\theta)=X^{fit}e^{-c\cdot \theta^n}$ of $Z_T$, $Z_{X_{2a}}$ and $Z_{X_{2b}}$ at two lattice spacings, and compared with the values $Z(\epsilon_0)$ with the highest available precision $\epsilon_0=2.4\times 10^{-14}$.}
\label{tab:fit_param}
\end{table}

The values of $Z^{fit}$, $c$ and $n$ of different operators using the lowest 4 gauge fixing precisions ($\theta\ge2.4\times 10^{-6}$ for a06m310 and $\theta\ge2.4\times 10^{-5}$ for a12m310) at two lattice spacings, are collected in Table~\ref{tab:fit_param}, and compared with those using the highest $\theta=2.4\times 10^{-14}$. Even with the volume source, the $c$ and $n$ for the scalar current using the clover fermion still has very large uncertainty and then is not shown here. 

Even with the volume source, the $c(Z_S)$ and $n(Z_S)$ for the scalar current using the clover fermion still has very large uncertainty and then is not shown here. 
The value \( c(Z_T) = -0.7(1) \) obtained with clover fermions differs from the overlap fermion result \( c(Z_T) = -0.47(2) \) on the identical a06m310 ensemble and renormalization scale \(\mu\). These discrepancy imply that the dependence on gauge-fixing precision could  depend on the fermion discretization. Based on the comparison of the results at two lattice spacings with the same $\mu$, $n$ is always around 0.5, while $|c|$ becomes larger at smaller lattice spacing, as we found in the previous subsection using the overlap fermion. 

As shown in Table~\ref{tab:fit_param}, the extrapolated values \(Z^{\text{fit}}\) agree with the results from high-precision gauge fixing to within at most 0.03\%. This deviation is smaller than the typical statistical uncertainty in hadron matrix elements. Compared to the value obtained at the highest precision used in the fit, the extrapolation reduces the deviation by at least an order of magnitude. Given \(n \sim 0.5\) in the empirical formula, this corresponds to an improvement of two orders of magnitude in the effective gauge-fixing precision \(\theta\).

\section{Applications on ${\xi}$ Gauge}\label{sec:xires}

The success of the precision extrapolation method in Landau gauge suggests that a similar approach would resolve precision issues in the $\xi$ gauge.
Unlike Landau gauge which can be fixed to the machine precision, the minimal attainable $\theta$ in the $\xi$ gauge is inherently limited by the current gauge fixing algorithm, especially when $\xi$ is large.

As the $\xi$-gauge dependence of $Z^{RI}_{S,T}$ is known perturbatively to 3 loops, comparing with non-perturbative determinations of $Z^{RI}_{S,T}(\xi)$ offers a powerful consistency check. This comparison tests both the sufficiency of $\xi$-gauge fixing precision and the validity of precision extrapolation approach.

\begin{table}
\begin{tabular}{c c c c c c c}
\hline
$\xi$           & 0.0 & 0.2 & 0.4 & 0.6 & 0.8 & 1.0 \\
\hline   
\# of $\theta$ & 21  & 11  & 9   & 8   & 7   & 6  \\
\hline
\end{tabular}
\caption{Number of gauge fixing precision  $\theta$ for different $\xi$. The largest $\theta$ is $10^{-2}$ and decreases progressively by a factor of $\sqrt{10}$, yielding a precision sequence like $\{10^{-2}, 10^{-2.5},10^{-3},...\}$.}
\label{tab:xiprec}
\end{table}

We execute our calculations on the ensemble a06m310 using the same valence overlap fermion point source propagators generated for the Landau gauge calculation, and do the precision extrapolation for all the combinations of $\xi$ and $\mu$. 
As Fig.~\ref{fig:fixing} shows, the increasing lower band of gauge fixing precision $\theta$ at larger $\xi$ reduces the number of available data points, as quantified in Table~\ref{tab:xiprec}. 

In principle, the $\xi$-gauge dependence of $Z^{\rm RI}(\xi,\mu;a)$ should match the perturbatively calculated result under dimensional regularization at 3-loop~\cite{Gracey:2003yr}, up to discretization errors:  
\begin{align}\label{eq:xi_ratio}
    R_X(\xi,\mu;a) &= \frac{Z_X^{\rm RI}(\xi,\mu;a)}{Z_X^{\rm RI}(0,\mu;a)} \frac{Z_X^{\rm RI,pert}(0,\mu)}{Z_X^{\rm RI,pert}(\xi,\mu)} \nonumber \\
    &= 1 + \mathcal{O}(a^2\mu^2).
\end{align}  
However, in practice, the value of $\xi$ in Eq.~(\ref{eq:xi_ratio}) is sensitive to the definition of the bare coupling $g_0$, as discussed earlier. 

\begin{figure}
    \centering
    \includegraphics[width=0.95\linewidth]{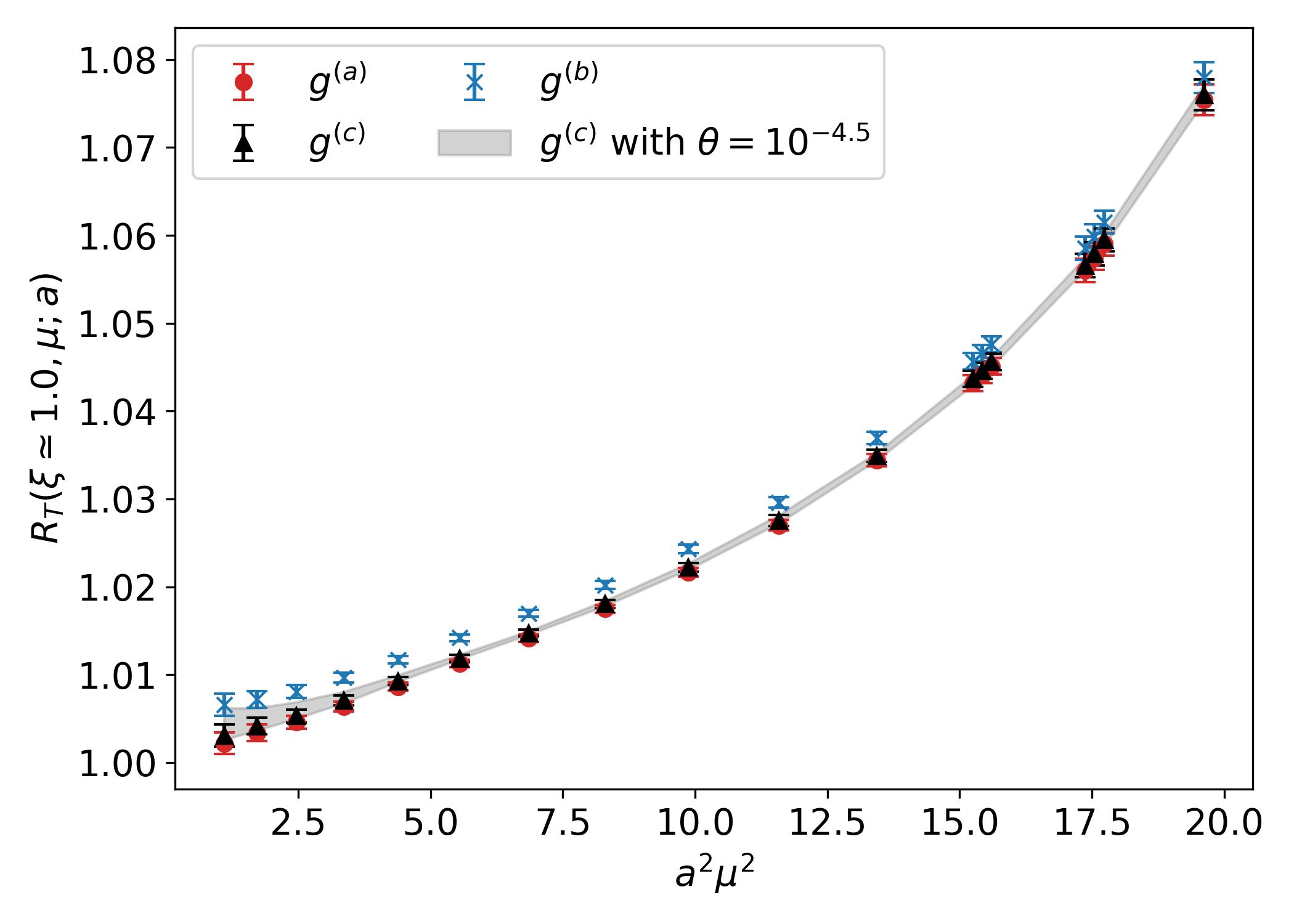}
    \caption{Ratio of $Z_T$ defined in Eq.~(\ref{eq:xi_ratio}) which should be unity up to the $a^2\mu^2$ error, with $\xi=(g^{(a)}_0/g_0)^2$ (in the $\theta\rightarrow 0$ limit) with the naive definition $g_0=g_0^{(a)}$ (red dots), tadpole improved one $g_0^{(b)}=1.13g_0^{(a)}$ (blue crosses), and also $u_0$ approximation $g_0^{(c)}=1.02g_0^{(a)}$ (black triangles). The gray band using the $g_0^{(c)}$ with $\theta=10^{-4.5}$ is also shown for comparison.}
    \label{fig:xi_1p0}
\end{figure}

In Fig.~\ref{fig:xi_1p0}, we plot the ratio $R_T((g^{(a)}_0/g_0)^2, a, \mu)$ of the tensor operator for three definitions of $g_0$ in the $\theta\rightarrow 0$ limit: 1) $g_0=g^{(a)}_0$ from the naive definition (red dots), 2) $g^{(b)}_0$ with full tadpole improvement (blue crosses), and 3) $g^{(c)}_0$ from the $u_0$ approximation (black triangles). We can see that the $a^2p^2$ extrapolated value using either $g_0^{(a)}$ or $g_0^{(c)}$, are closer to $1$ than that using $g_0^{(b)}$ and then can be considered as a good choice of $g_{0}$. We adopt $g^{(c)}_0$ to define the effective $\xi$ for two reasons: first, it can be extracted directly from gauge configurations without requiring knowledge of the discretized action; second, unlike the naive definition $g_0^{(a)}$, it implicitly incorporates the tadpole improvement necessary for a well-convergent perturbative series.

For comparison, we also show the $R_T((g^{(a)}_0/g^{(c)}_0)^2, a, \mu)$ with finite gauge fixing precision $\theta=10^{-4.5}$ as gray band for comparison. We observe that the precision-extrapolated values align with those obtained using $\theta = 10^{-4.5}$, albeit with slightly smaller statistical uncertainty. This suggests that systematic uncertainties arising from precision extrapolation are well-controlled in this case.

\begin{figure}
    \centering
    \includegraphics[width=0.95\linewidth]{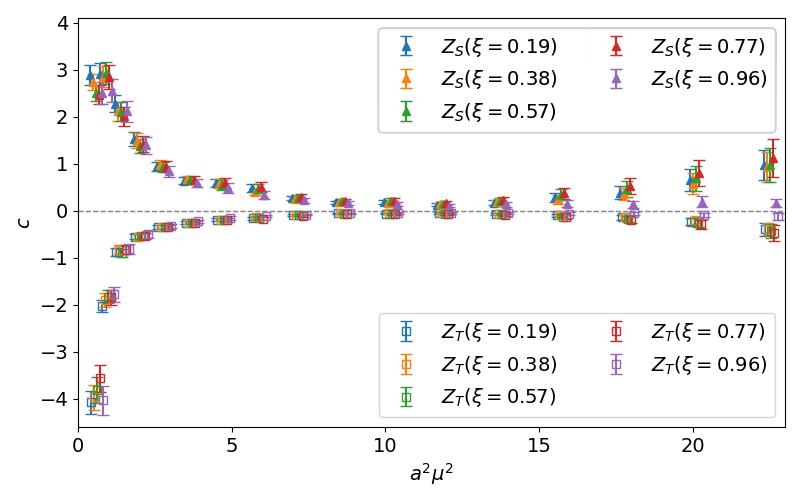}
    \includegraphics[width=0.95\linewidth]{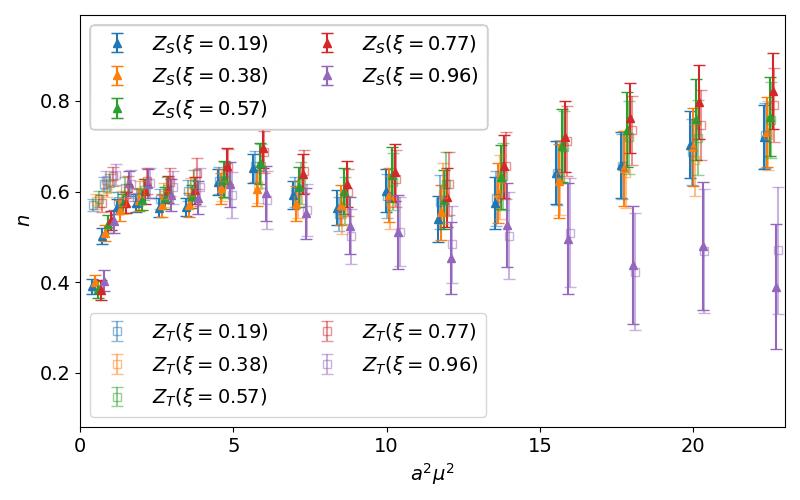}
    \caption{Gauge fixing precision parameters $c$ (upper panel) and $n$ (lower panel) for the ratio $X(\theta)/X(0)=e^{-c\cdot \theta^n}$ of $Z_S$ and $Z_T$ across various RI/MOM scale $\mu$ and gauge parameter $\xi$. 
    Results $c$ and $n$ of $\xi\neq0$ show good consistency with different $\xi$, while sensitive to $a^2\mu^2$ and current operator.}
    \label{fig:xi_cn}
\end{figure}

Fig.~\ref{fig:xi_cn} shows the fitting results of $c$ and $n$ using the empirical formula in Eq.~(\ref{eq:empi}), with good $\chi^2$/d.o.f. in all the cases. We can see that $n$ is also around 0.5 regardless of $\xi$ and $\mu$, while $|c|$ becomes larger at both ends of the $a^2\mu^2$ range. Since the empirical formula describes the data well for all values of $\xi$ and $a^2\mu^2$ we studied, the precision extrapolation is expected to reduce the gauge-fixing deviation. Based on the established results in Landau gauge, this suppression factor is typically $\sim 10^{2}$. Consequently, for the largest value $\xi \sim 1$, the method reduces the deviation from the practically achievable tolerance of \(\theta \sim 10^{-4.5}\) to an effective level of \(\theta \sim 10^{-6.5}\).

\begin{figure}
    \centering
    \includegraphics[width=0.95\linewidth]{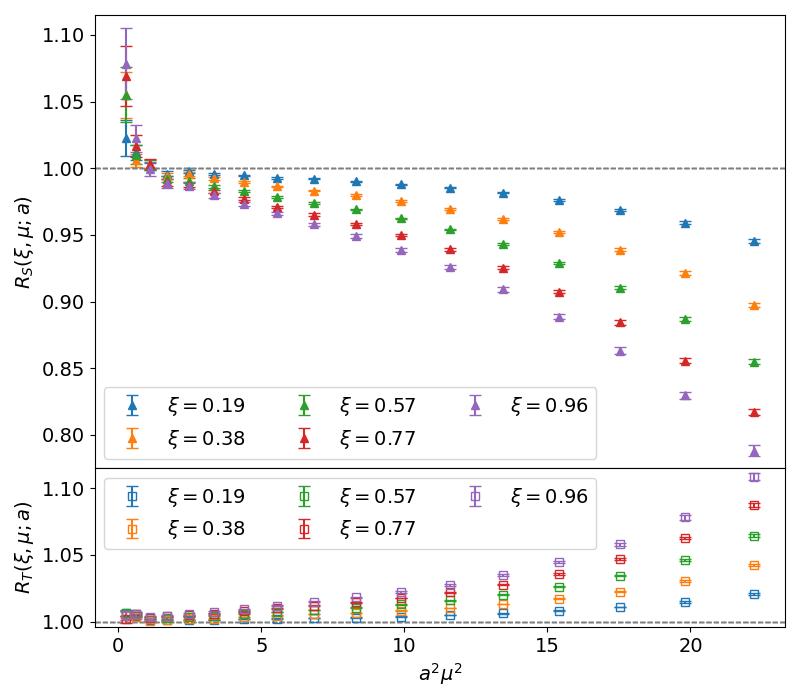}
    \caption{The ratios $R_S$ (upper panel) and $R_T$ (lower panel) as the function of $a^2\mu^2$ with different $\xi$.}
    \label{fig:xiaccu}
\end{figure}

The ratios $R_S$ (upper panel) and $R_T$ (lower panel) are shown in Fig.~\ref{fig:xiaccu} as the function of $a^2\mu^2$ with different $\xi$. Then we use the following polynomial ansatz to fit the data of the operator $X$ using the momentum range of $a^2p^2\in(2,20)$,
\begin{align} \label{eq:discerr}
R_{X}(\xi,\mu;a)=c_{0,X}(\xi)\left(1+\sum_{i=1}^3c_{i,X}(\xi)\left(a^2\mu^2\right)^i\right)
\end{align}
where $c_0-1$ represents the deviation of the numerical $\xi$ gauge fixing after the $a^2\mu^2\rightarrow 0$ extrapolation.

\begin{figure}
    \centering
    \includegraphics[width=0.95\linewidth]{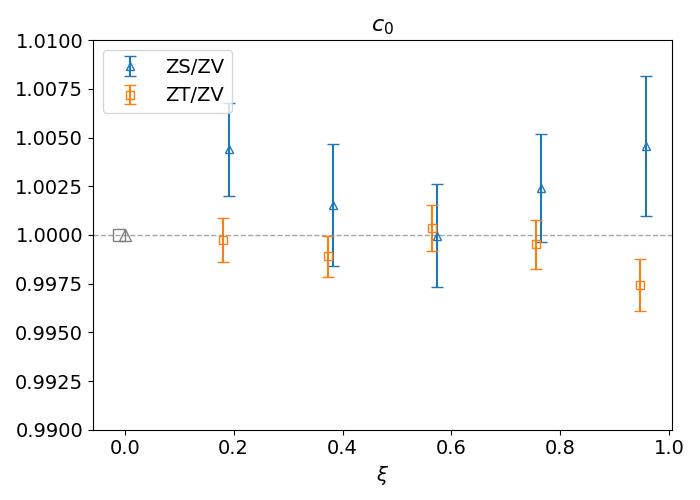}
    \caption{Extrapolated $R_S$ (blue triangles) and $R_T$ (orange boxes) after the $a^2\mu^2\rightarrow 0$ extrapolation, with different $\xi$.}
    \label{fig:xic0}
\end{figure}

As illustrated in Fig.~\ref{fig:xic0}, the fitted $c_0$ for different $\xi$'s are consistent with 1 up to 2$\sigma$ with no more than 0.3\% statistical uncertainty. All the fitting parameters obtained through full Jackknife resampled fitting are collected in Tab.~\ref{tab:xiextrapolation}, and we can see that $c_{1,S/T}$ can be described by

$c_{1,S}=-0.00803(51)\xi$ and $c_{1,T}=0.00326(24)\xi$ within the statistical uncertainty. 

\begin{figure}
    \centering
    \includegraphics[width=0.95\linewidth]{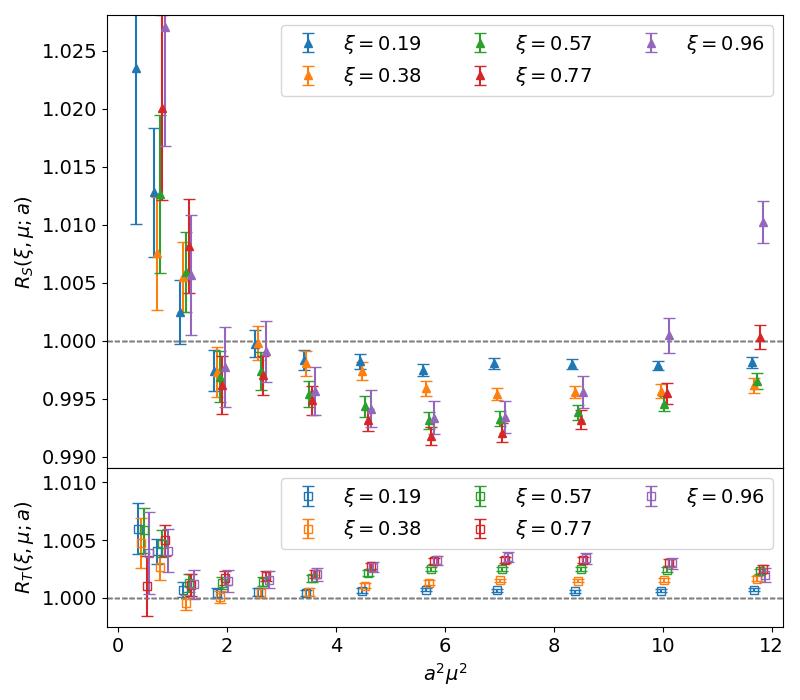}
    \caption{$R_S$ (upper panel) and $R_T$ (lower panel) for using $\tilde{\xi}=\xi a^2\mu^2/(4\mathrm{sin}^2(a\mu/2))$ in the perturbative matching.}
    \label{fig:xi_mod}
\end{figure}

While the Symanzik-improved gauge action is accurate to $\mathcal{O}(a^4)$, the renormalization constants for quark bilinear operators inherit an $\mathcal{O}(a^2)$ error from the fermion actions. Therefore, the overall deviation cannot be expected to be smaller than $\mathcal{O}(a^2)$. The $a^2\mu^2$ error in $R_X$ would originate from the discretized gauge-fixing condition in Eq.~(\ref{eq:condition}), which is equivalent to using a $\mu$-dependent $\xi$ parameter. By defining an effective $\tilde{\xi} = \xi  a^2\mu^2 / (4\sin^2(a\mu/2))=\xi(1+\frac{1}{12}a^2\mu^2+{\cal O}(a^4\mu^4))$ and using $\tilde{\xi}$ in the perturbative renormalization constant $Z_{X}^{\rm RI,pert}$ used in Eq.~(\ref{eq:xi_ratio}), we suppress the $a^2\mu^2$ dependence in $R_{X=S,T}$ to the 1\% level or less for $a^2\mu^2 \le 10$, as shown in Fig.~\ref{fig:xi_mod}. This result suggests that an improved gauge-fixing condition, like the one proposed in Ref.~\cite{Bonnet:1999mj}, would be highly effective in suppressing this discretization error.

\begin{table}
\begin{tabular}{c| c c c c c}
\hline
&$\xi$ & $c_0$       & $c_1(\times 10^{-3})$ & $c_2(\times 10^{-4})$ &  $c_3(\times 10^{-5})$ \\
\hline
& 0.191 & 1.0044(24) & -2.71(61) & 1.85(49) & -0.81(13) \\
& 0.383 & 1.0015(31) & -3.14(77) & 1.45(63) & -0.96(16) \\
$Z_S/Z_V$& 0.574 & 1.0000(27) & -4.23(61) & 1.71(54) & -1.22(15) \\
& 0.765 & 1.0024(28) & -6.21(70) & 2.28(66) & -1.45(20) \\
& 0.957 & 0.9973(25) & -6.12(74) & 1.25(69) & -1.24(19) \\
\hline
& 0.191 & 0.9997(11) & 0.57(32) & -0.40(27) & 0.24(07) \\
& 0.383 & 0.9989(11) & 1.37(29) & -0.96(26) & 0.54(07) \\
$Z_T/Z_V$& 0.574 & 1.0004(12) & 1.56(31) & -1.01(27) & 0.69(08) \\
& 0.765 & 0.9995(12) & 2.33(35) & -1.40(30) & 0.91(09) \\
& 0.957 & 0.9991(14) & 3.14(36) & -2.04(30) & 1.27(09) \\
\hline
\end{tabular}
\caption{Fitting parameters of Eq.~(\ref{eq:discerr}) for different $\xi$.}
\label{tab:xiextrapolation}
\end{table}

For illustration, we define the \(a^2\mu^2\)-extrapolated \(Z^{\rm RI}\) as  
\begin{align}
    \tilde{Z}^{\rm RI,latt}(\xi,\mu) \equiv \frac{Z^{\rm RI,latt}(\xi,\mu;a)}{f(a^2\mu^2,\xi)f_0(a^2\mu^2)},
\end{align}
where \(f(a^2\mu^2,\xi) = 1 + \sum_{i=1}^3 c_i(\xi) (a^2\mu^2)^i\) represents the discretization error obtained by fitting Eq.~(\ref{eq:discerr}), and \(f_0(x) \equiv 1 + \sum_{i=1}^3 d_i (a^2\mu^2)^i\) quantifies additional discretization errors in \(Z^{\overline{\mathrm{MS}}}(2~\mathrm{GeV})\) obtained though \(Z^{\rm RI,latt}(0,\mu;a)\) under Landau gauge. In specific, $f_0$ term is extracted through the polynomial fit of the following combination,  
\begin{align}
Z^{\rm RI,latt}(0,\mu;a) &\frac{Z^{\overline{\mathrm{MS}},\mathrm{pert}}(2~\mathrm{GeV})}{Z^{\mathrm{RI, pert}}(0,\mu)} \nonumber\\
&\quad \quad = Z^{\overline{\mathrm{MS}},\mathrm{latt}}(2~\mathrm{GeV}) f_1(a^2\mu^2),
\end{align}
with $Z^{\overline{\mathrm{MS}},\mathrm{latt}}(2~\mathrm{GeV})$ and $d_{i=1,2,3}$ as fit parameters.
Here, the perturbative ratio \(\frac{Z^{\overline{\mathrm{MS}},\mathrm{pert}}(2~\mathrm{GeV})}{Z^{\mathrm{RI, pert}}(0,\mu)}\) is derived by matching \(Z^{\mathrm{RI, pert}}\) to \(Z^{\overline{\mathrm{MS}}}\) at scale \(\mu\) first, then evolving to 2 GeV using renormalization group equations. We further define the perturbative \(\xi\)- and \(\mu\)-dependence of \(Z^{\mathrm{RI, pert}}(\xi,\mu)\) as  
\begin{align}
&\tilde{Z}^{\rm RI, pert}(\xi,\mu) \equiv \frac{Z^{\mathrm{RI, pert}}(\xi,\mu)}{Z^{\overline{\mathrm{MS}},\mathrm{pert}}(2~\mathrm{GeV})} Z^{\overline{\mathrm{MS}},\mathrm{latt}}(2~\mathrm{GeV})\nonumber\\
&=\frac{Z^{\mathrm{RI, pert}}(\xi,\mu)}{Z^{\overline{\mathrm{MS}},\mathrm{pert}}(\mu)}
\frac{Z^{\overline{\mathrm{MS}},\mathrm{pert}}(\mu)}{Z^{\overline{\mathrm{MS}},\mathrm{pert}}(2~\mathrm{GeV})} Z^{\overline{\mathrm{MS}},\mathrm{latt}}(2~\mathrm{GeV}),
\end{align} 
where both the ratios $\frac{Z^{\mathrm{RI, pert}}(\xi,\mu)}{Z^{\overline{\mathrm{MS}},\mathrm{pert}}(\mu)}$ and $\frac{Z^{\overline{\mathrm{MS}},\mathrm{pert}}(\mu)}{Z^{\overline{\mathrm{MS}},\mathrm{pert}}(2~\mathrm{GeV})}$ are finite and can be obtained by the perturbative calculations~\cite{Gracey:2003yr}.

\begin{figure}
    \centering
    \includegraphics[width=0.95\linewidth]{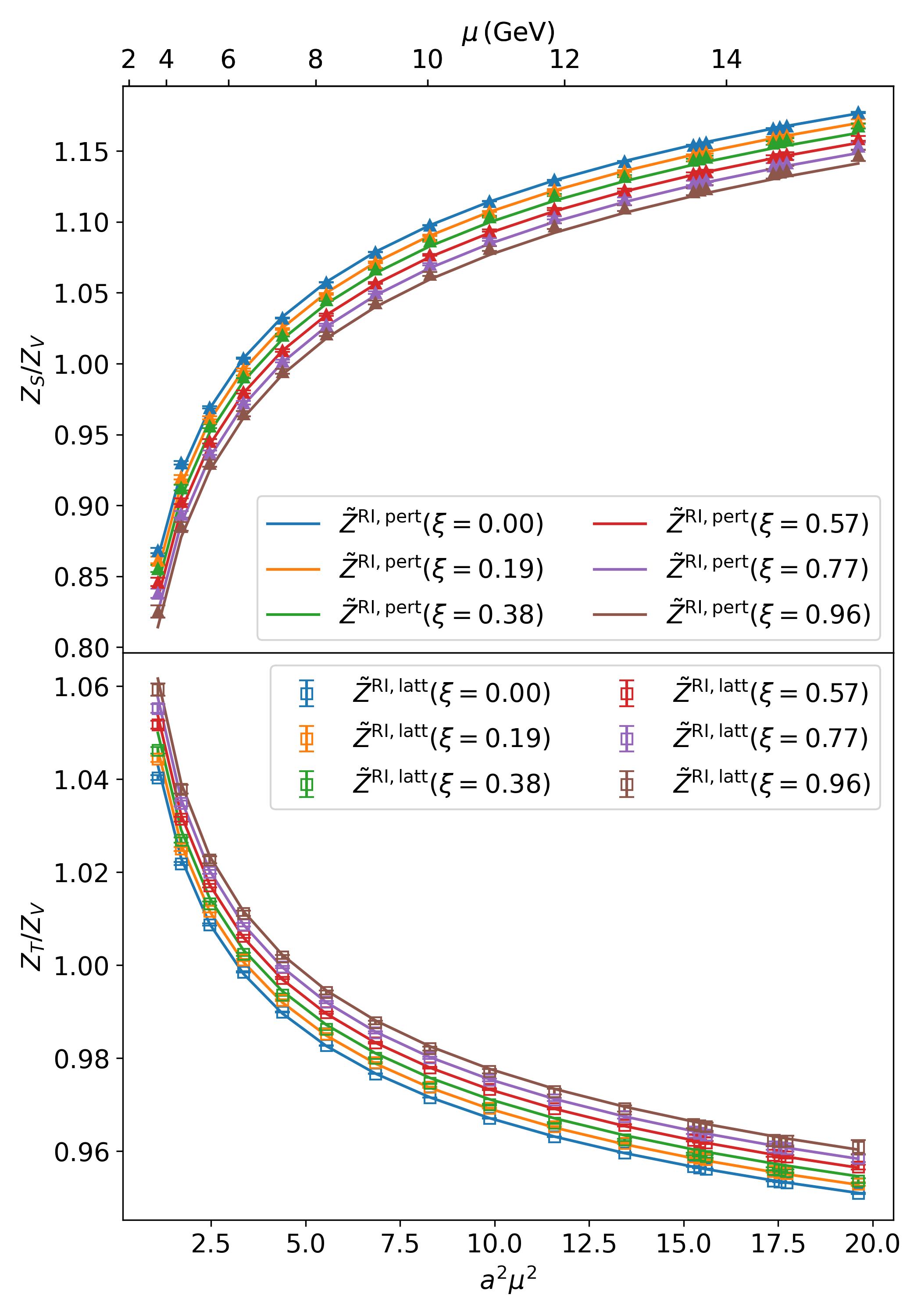}
    \caption{Comparison of renormalization constants under RI/MOM regenerated from perturbation theory with that of lattice computation. Lines are perturbation results and data points are from lattice. The consistency of the line and points exhibits the success of taking $\xi$ dependence and discretization error apart. }\label{fig:z_compare}
\end{figure}

Figure~\ref{fig:z_compare} presents a comparison between the extrapolated lattice results $\tilde{Z}^{\rm RI,latt}(\xi,\mu)$ (colored data points) and their perturbative counterparts $\tilde{Z}^{\rm RI,pert}(\xi,\mu)$ (colored lines) for both scalar (upper panel) and tensor (lower panel) quark bilinear operators. The observed agreement primarily demonstrates the effectiveness of the $a^2\mu^2$ polynomial in describing the discrepancy between lattice computations and perturbative calculations. A more rigorous comparison would require continuum extrapolation using lattice data at multiple spacing values.

\section{Summary}\label{sec:sum}

In this work, we first establishes the empirical dependence of RI/MOM renormalization constants for quark bilinear operators on Landau gauge-fixing precision. We demonstrate that the deviation scales universally as \(\sim \theta^{1/2}\) across operators (scalar, tensor, EMT), RI/MOM scales, and fermion discretizations.

Thus imprecise Landau gauge fixing differs from \(\xi\) gauge fixing with \(\xi \propto \theta\), owing to the distinct parameter dependencies of physical quantities (\(\sim \theta^{1/2}\) versus \(\xi\)), which originate from the respective distributions of deviations from the Landau gauge.
It further allows us to rule out an interpretation of imprecise gauge fixing within the soft covariant gauge framework~\cite{Fachin:1991pu,Henty:1996kv} based on the continuum quantization proposal~\cite{Parrinello:1990pm,Zwanziger:1990tn}. Perturbatively, the soft covariant gauge approaches the Landau gauge in the \(M^2 \to \infty\) limit with a \(1/M^2\) correction~\cite{Fachin:1993qg}. If our imprecise gauge fixing were equivalent to a soft covariant gauge with some \(M^2\), we would expect deviations scaling as \(1/M^2\). In contrast, our observed scaling is \(\sim \theta^{1/2} \propto 1/M\), which is qualitatively different. Moreover, as noted in the literature~\cite{Henty:1996kv}, reaching the \(1/M^2 \to 0\) limit in the soft covariant gauge corresponds to a weighted average of different minimum of $\theta$ and then numerically challenging. Thus it is unlikely to be achieved by imprecise gauge fixing for a given minimum of $\theta$.

Building on this analysis, we introduce a precision extrapolation procedure that eliminates gauge-fixing residuals in the \(\xi\) gauge. Applying this method, our lattice calculations of \(Z^{\rm RI}_{S,T}\) for \(\xi \le 1\) achieve 0.3\% consistency with three-loop perturbative results.

Nevertheless, the empirical form for the \(\theta\) dependence of the $\xi$-dependent quantities remain an assumption. Unlike Landau gauge, the \(\xi\)-gauge fixing problem is ill-posed, so the precision extrapolation here should be viewed as a practical tool rather than a rigorously justified limit.

Given that precision sensitivity grow significantly at small lattice spacings \(a\) or low RI/MOM scales \(\mu\) (e.g., \(\mu \lesssim 1\) GeV), this extrapolation method is especially valuable for accurate infrared parton studies. It also paves the way for systematic investigations of \(\xi\)-dependent quark and gluon propagators and their non-perturbative infrared interactions under controlled gauge-fixing uncertainties. As such, this work provides a foundation for improving a range of phenomenological models of non-perturbative QCD.

Furthermore, improved gauge fixing conditions~\cite{Bonnet:1999mj} could significantly suppress the \(\xi\)-dependent \(\mathcal{O}(a^2\mu^2)\) discretization errors observed in Fig.~\ref{fig:xiaccu}, making this an important direction for future study. At the same time, the precision extrapolation method becomes unreliable for \(\xi \gtrsim 1.2\), as the minimal achievable residual \(\theta_{\rm min} \sim 10^{2.9\xi-7.3}\) drastically reduces the number of viable data points. Extending these calculations to larger \(\xi\) values will therefore require the development of more sophisticated gauge-fixing algorithms. 

\section*{Acknowledgment}
We thank MILC Collaboration for providing their HISQ gauge configuration, and Ying Chen for valuable comments and suggestions.
The calculations were performed using the Chroma software suite~\cite{Edwards:2004sx} with QUDA~\cite{Clark:2009wm,Babich:2011np,Clark:2016rdz} and GWU code~\cite{Alexandru:2011ee,Alexandru:2011sc} through HIP programming model~\cite{Bi:2020wpt}. The numerical calculations were carried out on the ORISE Supercomputer, HPC Cluster of ITP-CAS and Advanced Computing East China Sub-center. This work is supported in part by National Key R\&D Program of China No.2024YFE0109800, NSFC grants No. 12293060, 12293062, 12435002, 12447101, 12447102 and 12175030, the science and education integration young faculty project of University of Chinese Academy of Sciences, the Strategic Priority Research Program of Chinese Academy of Sciences, Grant No. YSBR-101.

\bibliographystyle{apsrev4-1}
\bibliography{ref}

\end{document}